\documentclass[a4paper,titlepage]{article}
\pdfoutput=1
\usepackage{graphics,natbib}
\usepackage{amssymb,bm}

\newcommand{\vect}[1]{#1}
\newcommand{\Var}{{\rm Var}}
\newcommand{\E}{{\rm E}}

\begin{document}

\title{Bayes linear kinematics in a dynamic Bayesian survival model}
\author{Kevin J.\ Wilson\footnote{Email: {\tt kevin.wilson@newcastle.ac.uk}} and 
Malcolm Farrow\footnote{Email: {\tt malcolm.farrow@newcastle.ac.uk}} \\
School of Mathematics and Statistics, Newcastle University, UK.}
\maketitle

\begin{abstract}
Bayes linear kinematics and Bayes linear Bayes graphical models provide an extension of Bayes linear methods so that full conditional updates may be combined with Bayes linear belief adjustment. In this paper we investigate the application of this approach to survival analysis with time-dependent covariate effects, a more complicated problem than previous applications. We use a piecewise-constant hazard function with a prior in which covariate effects are correlated over time.  The need for computationally intensive methods is avoided and the relatively simple structure facilitates 
interpretation. Our approach eliminates the problem of non-commutativity which was observed in earlier work by Gamerman. We apply the technique to data on survival times for leukemia patients.

\vspace{1cm}

{\bf Keywords}: Bayes linear kinematics, Bayes linear Bayes graphical model, dynamic model, piecewise constant hazard, survival analysis, time-dependent covariate effects
\end{abstract}

\section{Introduction}
A Bayes linear analysis \citep{Gol07} differs from a full Bayesian analysis in that only first and second order moments are specified in the prior. Posterior (termed {\em adjusted}) moments are then calculated when data are observed. The introduction of Bayes linear kinematics and Bayes linear Bayes models \citep{Gol04} extends Bayes linear methods to allow the incorporation of observations of types which are not readily accommodated in a straightforward Bayes linear analysis. For example, beliefs about certain unknown quantities might be updated by full conditional Bayesian inference when observations are made on conditionally Poisson or binomial variables and then information can be propagated between these unknowns, or to other unknowns, via a Bayes linear belief structure. This approach avoids the need for computationally intensive methods such as Markov chain Monte Carlo which are often required in standard Bayesian analyses. 
Computational time in evaluating posterior distributions can be an important issue in areas such as design of experiments \citep{Muller04}, clinical decision rules or evaluation of diagnostic tests. Such analyses may require the repeated evaluation of posterior distributions given large numbers of simulated data sets. In such cases, the Bayes linear kinematic method may provide an effective emulator \citep{Jones16}.

\cite{Wils10} introduced the use of a link function to map the range of an unknown, such as the mean of a Poisson distribution, onto the whole real line and improve the linearity of the relationships represented by the Bayes linear structure. In this paper we show how Bayes linear kinematics may be applied to a more complicated problem, specifically in the analysis of survival data, and that this brings appealing advantages over standard techniques. Our analysis uses death and censoring times, in contrast to the relatively simple actuarial methods developed in \cite{Wils10}. We use a piecewise constant hazards model with temporally-dependent hazard priors. We combine fully Bayesian conjugate updating for individuals in intervals and Bayes linear kinematic updating to propagate changes in belief to other individuals and intervals. 
Our model is related to that of \cite{Gam91} but, using the Bayes linear kinematic approach, we overcome the problem of non-commutativity of updates observed by Gamerman. 

We consider Bayesian analysis from a subjectivist perspective \citep{Gol06,Lin06}. Therefore we give attention to the appropriate specification of prior beliefs.

The remainder of the paper is structured as follows. Section \ref{sec:sur} gives an overview of proportional hazards models and the piecewise constant hazards model and reviews the model of \cite{Gam91}. 
In Section \ref{sec:bl} we give a brief introduction to the results of \cite{Gol04}. In Section \ref{sec:blk} we describe our Bayes linear kinematic solution to the survival problem in four stages; the guide relationship, system evolution, use of Bayes linear kinematics and calculation of the expectations and variances. The usefulness of the approach is illustrated with an example involving survival times of leukemia patients in the North-West of England in Section \ref{sec:ex}. Some conclusions and areas for further work are presented in Section \ref{sec:con}.

\section{Survival analysis}
\label{sec:sur}
\subsection{Introduction}
In this paper we investigate the application of Bayes linear kinematics and Bayes linear Bayes models in survival analysis, specifically a proportional hazards model with piecewise constant hazards. Survival analysis is concerned with modelling the time elapsed, known as the survival time, until some event occurs.  For convenience we shall refer to the event as ``death''.

The survival time $t$ of an individual is a realisation of a random variable $T$. Associated with $T$ is a survival function
$ S(t)=\Pr(T\geq t), $ a probability density function $ f(t) $ and a hazard function $ h(t)=f(t)/S(t). $
Censoring of observations is a common feature of survival data. In right censoring all that is known is that $t>c$ for some value $c$, in left censoring this condition is $t<c$ and in interval censoring $c_{1}<t<c_{2}$, for some values $c_{1},c_{2}$. In this paper we  consider only right censoring, which is the most common type, and assume that the censoring is non-informative. That is, the survival time $T$ is independent of the mechanism which causes an observation to be censored. Further information on Bayesian survival analysis can be found in \cite{Kle97} and \cite{Ibr01}.

\subsection{Proportional hazards models}
Suppose we have individuals $i=1,\ldots,p$ and individual $i$ has covariate values $    x_{i}=(x_{i,0},x_{i,1},\ldots,$ $x_{i,q})'$ where, typically, $x_{i,0}\equiv 1$. Associated with  individual $i$ is a hazard function $h_{i}(t)$.  A popular and appealing way to relate the covariate values to the survival distribution for an individual is to make the proportional hazards assumption \citep{Cox72}. Then we can write
$h_{i}(t)=\phi_{i}h_{0}(t),$
where $\phi_{i}$ is a constant with respect to time and $h_{0}(t)$ is a baseline hazard function. We can relate an individual's hazard function to $x_{i}$, the individual's covariate vector, by setting
\begin{equation}
\phi_{i}=\exp\left(\sum_{k=1}^{q} x_{i,k}\beta_{k}\right),
\label{static}
\end{equation}
for some parameters $ \beta_{1}, \ldots , \beta_{q}$, which, in a simple proportional hazards model, remain constant over time. 

\subsection{Piecewise constant hazards model}
We might be unwilling to assume a particular form for the baseline hazard function $ h_{0}(t)$. A simple and much investigated way to relax this assumption is to use a piecewise constant hazards 
model \citep[eg.][]{Ibr01}. Time is partitioned into disjoint intervals. In each interval a constant hazard is specified 
but the hazards are allowed to vary from interval to interval.

Furthermore we may wish to allow the effects of the covariates, represented by the coefficients $ \beta_{1}, \ldots , \beta_{q}$, to vary from one time interval to another. This has led to the development of dynamic survival models in which the coefficients can vary over time \citep{Mar06}. We shall consider a dynamic model
$ h_{i}(t)=\exp\{    x_{i}'   \beta(t)\}, $
where $     x_{i}' = (1,x_{i,1}, \ldots , x_{i,q}) $ and $     \beta'(t) = (\beta_{0}(t), \beta_{1}(t), \ldots , \beta_{q}(t)) $ with $ \beta_{0}(t) = \log\{ h_{0}(t)\}, $
so that we can model changes in the effects of the covariates over time. The static model in (\ref{static}) is then a special case of this more general model.

We choose fixed time points $\tau_{0},\tau_{1},\ldots,\tau_{r}$ such that $\tau_{0}=0$ and $ \tau_{r} \rightarrow \infty$.
This partitions time into intervals. We say that the $j^{\rm th} $ interval is $R_{j}=[\tau_{j-1},\tau_{j})$. Then, for $\tau_{j-1}\leq t < \tau_{j}$, the baseline hazard is $h_{0}(t)=\lambda_{0,j}$ and the hazard function for individual $i$ is
$h_{i}(t)=\lambda_{i,j}=\phi_{i,j}\lambda_{0,j}=\exp(  \eta_{i,j}),$
where $   \eta_{i,j}=   x_{i}'   \beta_{j} $ is the {\em linear predictor} and $  \beta_{j}=(\beta_{j,0}, \ldots , \beta_{j,q})'. $
That is, the hazard for each individual remains constant through each of the time intervals. The integrated hazard $H_{i}(t)= \int_{0}^{t}h_{i}(u)du$ is then
\begin{eqnarray*}
H_{i}(t)
& = & \sum_{k:\tau_{k}<t}\lambda_{i,k}(\tau_{k}-\tau_{k-1})+\lambda_{i,j}(t-\tau_{j-1}),
\end{eqnarray*}
for $k=1,\ldots,j-1$. 

If we condition on  $T\geq\tau_{j}$ then we obtain the conditional survival function and conditional probability density function for individual $i$ at time $t$. These are, for $ \tau_{j-1} \leq t < \tau_{j}, $
\begin{equation}
f_{i}(t\mid T\geq\tau_{j-1})=\lambda_{i,j}\exp\{-\lambda_{i,j}(t-\tau_{j-1})\},
\label{density}
\end{equation}
and
\begin{equation}
S_{i}(t\mid T\geq\tau_{j-1})=\exp\{-\lambda_{i,j}(t-\tau_{j-1})\}.
\label{survival}
\end{equation}
Thus the conditional density takes the form of a shifted exponential distribution.

In our prior distribution, the coefficients could be independent between time intervals \citep{Kalb78,Ibr01}. However, it would seem sensible that hazards in intervals which 
are close together are likely to be similar and so it would be beneficial to use a prior distribution in which hazard parameters in neighbouring intervals are correlated. 
Such a temporally-dependent prior has the effect, if sufficiently large 
correlations between intervals are used, of both smoothing the resulting posterior mean of the hazard function and providing extra information for later time periods in which there will be fewer individuals left in the study. Examples include \cite{SCG99} and \cite{Sargent97}. In the latter case the interval boundaries are the observed event times. See also \cite{SinhaDey97}. 

Bayesian methods using fully-specified priors, and requiring intense computation, include those proposed by \cite{Mck00} and \cite{Kim07}. 
McKeague and Tighiouart introduced a method, with temporally-dependent priors using Markov random fields, which gave a prior distribution to the interval boundaries based on a non-homogeneous Poisson process.  Updating was fully Bayesian and used the Metropolis-Hastings-Green algorithm. \cite{Kim07} considered a temporally-dependent proportional hazards model in the context of cure fractions. Their method allowed a random number of intervals and random interval lengths. They used noninformative priors and utilised reversible jump Markov chain Monte Carlo (MCMC) methods for updating.

\subsection{The Gamerman model}
\label{sec:gam}
The temporally-dependent prior for the coefficients  $\{ \beta_{j,k} \}$ of the linear predictor of the hazard function in successive intervals may be built using  a dynamic linear system evolution as in a dynamic linear model. \cite{West85} introduced dynamic generalised linear models for non-Gaussian time series.
\cite{Gam91} applied this idea to a piecewise constant hazards model as follows.

\begin{enumerate}
 \item  Each $\lambda_{i,j}$ was given a gamma prior distribution which was conjugate to the conditional likelihood for that individual in that interval. Within each interval, beliefs about individual hazards could then be updated straightforwardly using full Bayesian conditioning.
 \item An evolving system vector, as in a dynamic linear model, was used to specify the prior distribution for the coefficients $\{ \beta_{j,k}\}$. The joint distribution of the system vector at times $\tau_{0},\ldots,\tau_{r}$ was not fully specified, just the first and second order moments. The vector $\beta_{j}$ for interval $j$ was given a mean vector $    m_{j}$ and a covariance matrix $C_{j}$. The covariances between intervals were specified using the system evolution.
We call such a partial specification a second order belief  specification.
 \item A {\em guide relationship} between the parameters of the conjugate prior and the corresponding elements of the system vector was specified.
\end{enumerate}

Within each interval, beliefs about $ \lambda_{i,j} $ were updated when data were observed for that interval. This change in belief was then transmitted to an associated quantity $ \eta_{i,j} $  using the guide relationship, denoted here by $\approx$, as follows,
\begin{displaymath}
\log(\lambda_{i,j}) \approx \eta_{i,j} =    x_{i}'   \beta_{j}.
\end{displaymath}
As a result of the partial specification, the effects of changes in beliefs about $ \eta_{i,j}$ were then propagated to other individuals' hazards in the interval and to other intervals using Bayes linear updating rules.
 Thus inference combined both fully Bayesian and Bayes linear updating.   The structure led naturally to the use of a forward-filtering and backward-smoothing algorithm to compute posterior moments of the system vectors.

Gamerman found, however, that the calculated adjusted beliefs depended on the order in which data were included. `The dependence on the order that the observations are processed is of concern...The results, however, do not differ by much'. Such non-commutativity is clearly a cause for concern. 
A fully Bayesian analysis with a fully specified prior distribution, for example a multivariate normal distribution in place of the linear Bayes structure, will, of course, give commutativity, but at the expense of greatly increased computation and the necessity of a full distributional specification for the prior. There is a wide range of problems where the idea of combining nonlinear, fully specified, Bayesian updates for separate parts of a model with a linear Bayes structure to connect the parts, in a Bayes linear Bayes graphical model, is attractive. Subject to certain conditions, Bayes linear kinematics \citep{Gol04,Wils10} provides a method for commutative inference in such structures.

\section{Bayes linear kinematics}
\label{sec:bl}
In a traditional Bayesian analysis a full joint prior distribution is specified for all observables and unknown quantities such as parameters. Prior beliefs are then updated, by conditioning on the observations and using Bayes theorem, and posterior distributions are calculated.

A Bayes linear analysis \citep{Gol07} differs from a full Bayesian analysis in that only first and second order moments are specified in the prior. Posterior (termed {\em adjusted}) moments are then calculated. Consider two vector random quantities $  A=(a_{1},\ldots, a_{n_{A}})'$ and $   B=(b_{1},\ldots,b_{n_{B}})'$.  Suppose that a full second order prior specification has been made for the set $    A \cup B,$ in which the prior expectations are $ {\rm E}_{0}(A) = E_{0,A} $ and $ {\rm E}_{0}(B) = E_{0,B} $ and the prior variances and covariances are $ {\rm Var}_{0}(A) = V_{0,A,A},\ {\rm Var}_{0}(B) = V_{0,B,B} $ and $ {\rm Cov}_{0}(A,B) = V_{0,A,B}=V^{\prime}_{0,B,A}.$ 

Suppose that we will observe $ A. $ Bayes linear methods offer a procedure by which beliefs about $ B $ are updated by a process of linear fitting on $ A$. To do this, we minimise
$ {\rm E}_{0}\left\{(B - C A^{*})'(B - C A^{*})\right\}, $
with respect to the matrix $ C $, where $A^{*}=(1, a_{1},\ldots,a_{n_{A}})'$ and ${\rm E}_{0}$ denotes prior expectation. This gives the Bayes linear updating equations for the adjusted expectation and variance of $ B $ given $ A$:
\begin{equation}
\begin{array}{rcl}
{\rm E}( B \mid  A) & = & E_{0,B} + V_{0,B,A}V_{0,A,A}^{-1}( A - E_{0,A}), \\
{\rm Var}( B \mid  A) & = & V_{0,B,B} - V_{0,B,A}V_{0,A,A}^{-1}V_{0,A,B},
\end{array} \label{eq:Bayeslin}
\end{equation}
when $ V_{0,A,A} $ is invertible. When this matrix is not invertible a suitable generalised inverse such as the Moore-Penrose inverse can be used. 

Notice that an alternative representation of the same prior and adjusted beliefs is obtained by writing
\begin{equation}
B = E_{0,B} + M_{B|A}(A-E_{0,A}) + U_{B|A} \label{eq:one}
\end{equation}
where $ M_{B|A} = V_{0,B,A}V_{0,A,A}^{-1} $ and the random vector $ U_{B|A} $ has zero expectation and its variance is $ V_{0,B,B} - V_{0,B,A}V_{0,A,A}^{-1}V_{0,A,B}. $

Bayes linear kinematics \citep{Gol04}, named after the probability kinematics of \cite{Jef65}, deals with the case where, instead of observing the value of an unknown, such as $ A, $ in the Bayes linear structure, we receive information which causes us to change our beliefs about the unknown. Suppose that we receive information $ D_{A} $ which leads us to revise our moments for $ A $ to $  {\rm E}_{1}(A \mid D_{A}) = E_{1,A} $ and $ {\rm Var}_{1}(A \mid D_{A}) = V_{1,A,A}. $ The Bayes linear kinematic update of our beliefs about $ A \cup B $ is obtained by assuming that (\ref{eq:one}) continues to hold so that our adjusted beliefs for $ B $ become

\begin{eqnarray}
{\rm E}_{1}(B \mid D_{A}) & = & E_{0,B} + V_{0,B,A}V_{0,A,A}^{-1}( E_{1,A} - E_{0,A}), \label{eq:e1}  \\
{\rm Var}_{1}(B \mid D_{A}) & = & V_{0,B,A} V_{0,A,A}^{-1} V_{1,A,A} V_{0,A,A}^{-1}V_{0,A,B} \nonumber \\
                            &   & + V_{0,B,B} - V_{0,B,A} V_{0,A,A}^{-1}V_{0,A,B} . \label{eq:var1}
\end{eqnarray}

Consider the following simple example. We wish to perform a survival analysis for two groups of patients. Suppose that within each group the survival times of patients follow an exponential distribution with a common hazard rate and let this rate have a gamma prior distribution. Further suppose that the two hazard rates are correlated. That is, learning about the hazard rate in one group informs us about the hazard rate in the other group. Then, when we observe the survival times of a sample of patients from Group 1, this will allow us to update the hazard rate of that group using Bayes Theorem. Equations (\ref{eq:e1}) and (\ref{eq:var1}) then allow us to propagate this update through to the hazard rate in Group 2.   

Now consider the case where we have $J$ sets of random quantities $  \mathcal{X}_{1}, \ldots , \mathcal{X}_{J},$ where, for $ j=1, \ldots , J, $ the elements of $ \mathcal{X}_{j} $ are arranged as a vector $    X_{j}=(X_{j,1},\ldots,X_{j,n_{j}})'.$  The sets $\mathcal{X}_j$ need not be disjoint.
Suppose that a full second order prior specification has been made for $    \mathcal{X}=  \mathcal{X}_{1}\cup\ldots\cup    \mathcal{X}_{J}$ and that the elements of $ \mathcal{X} $ are arranged in a vector $ X. $ We denote this prior specification $S_{0}(X)=[{\rm E}_{0}(X), {\rm Var}_{0}(X)].$ Let the adjustment implied by (\ref{eq:Bayeslin}), if $ X_{k} $ is observed, be denoted $ {\rm E}_{(0)}(X \mid X_{k}) $ for the expectation and  $ \textrm{Var}_{(0)}(X \mid X_{k}) $ for the variance.  Now suppose that information $ D_{k} $ is received which causes the beliefs about $ X_{k}$ to be updated to $S_{1}(X_{k}\mid D_{k})= [ {\rm E}_{1}(X_{k}), {\rm Var}_{1}(X_{k})] $.

 Suppose that, in this new situation, the adjustment, given an observation of $ X_{k}, $ would be $ {\rm E}_{(1)}( X \mid X_{k}) $ for the expectation and $ \textrm{Var}_{(1)}( X \mid X_{k}) $ for the variance. Then the Bayes linear kinematic update for $ X$ is found by setting
\begin{eqnarray*}
{\rm E}_{(0)}( X \mid X_{k})={\rm E}_{(1)}( X \mid X_{k}), & \textrm{Var}_{(0)}( X \mid X_{k})=\textrm{Var}_{(1)}( X \mid X_{k}),
\end{eqnarray*}
that is, setting the adjusted expectations and variances using specifications $S_{0}(X)$ and $S_{1}(X\mid D_{k})$ equal to each other. Specifically we can use (\ref{eq:one}) with $ A $ replaced by $ X_{k} $ and $ B $ replaced by $ X. $ 

Now suppose that, for each $j$ $(j=1,\ldots,J)$, information $D_{j}$ is received once and beliefs are changed  for  $    X_{j}$ . A Bayes linear kinematic update can be made for $  X $ each time. However, successive Bayes linear kinematic updates are not necessarily commutative. Once (\ref{eq:e1}) and (\ref{eq:var1}) have been used for an update given $D_{1}$, the moments of the Bayes linear structure are changed and so the update by $D_{2}$ is changed. The Bayes linear kinematic method depends on the assumption that the updating formulae do not change, for example that (\ref{eq:one}) continues to hold. By straightforward repeated application of (\ref{eq:e1}) and (\ref{eq:var1}) we violate this assumption and it turns out that commutativity does not hold.  It is necessary to define Bayes linear kinematic formulae for updating by the whole of the data, based on an assumption analogous to the assumption that (\ref{eq:one}) continues to hold, but which will apply commutatively to intermediate steps, whatever the order in which we enter the data. 
\cite{Gol04} derived the conditions under which a commutative Bayes linear kinematic update exists and under which this update is unique. When a unique commutative update exists, it is given by 
\begin{eqnarray}
{\rm P}(X \mid D) & = & \sum_{j=1}^{J} {\rm P}(X \mid D_{j}) - (J-1) {\rm P}(X), \label{eq:ten} \\
{\rm P}(X \mid D){\rm E}(X \mid D) & = & \sum_{j=1}^{J} {\rm P}(X \mid D_{j}) {\rm E}(X \mid D_{j}) - (J-1) {\rm P}(X) {\rm E}(X), \label{eq:eleven}
\end{eqnarray}
where $ D = (D_{1}, \ldots , D_{J})$ and $ {\rm P}(X)= {\rm Var}(X)^{-1}$ is the precision matrix.

Goldstein and Shaw give formal proofs in the general case. We derive the results which we need for our purposes in the appendix.

\cite{Gol04} introduced a structure which they called a Bayes linear Bayes graphical model. This is constructed as follows. We have a collection $ \mathcal{X} $ of unknowns which are given a full second-order prior specification so that Bayes linear belief adjustments may be made. There are certain subsets $ \mathcal{X}_{1}, \ldots , \mathcal{X}_{J} $ of $ \mathcal{X} $ where, for each $ j, $ the subset $ \mathcal{X}_{j} $ is also part of a model which contains other quantities $ \mathcal{Z}_{j}. $ A full joint probabilistic prior specification is made over $ \mathcal{X}_{j} \cup \mathcal{Z}_{j}. $ The sets $ \mathcal{Z}_{1}, \ldots , \mathcal{Z}_{J} $ are disjoint and every element of $ \mathcal{Z}_{j} $ is conditionally independent of every element of $ \mathcal{X} \setminus \mathcal{X}_{j} $ and of every element of $ \mathcal{Z}_{k},\ k \neq j, $ given $ \mathcal{X}_{j}. $ Observing elements of $ \mathcal{Z}_{j} $ will cause us to revise our beliefs about $ \mathcal{X}_{j} $ and such changes can then be propagated to other elements of $ \mathcal{X} $ using (\ref{eq:ten}) and (\ref{eq:eleven}).

\subsection{Relevance to survival analysis}

The relevance of this to our survival analysis is that we regard the survival model, with second order prior specification for the hazard parameters, as a Bayes linear Bayes graphical model. Information is received, by observation of the deaths or survival of individuals in the time intervals, which changes our beliefs about components of the underlying Bayes linear structure.

\cite{Gam91} simply used equations (\ref{eq:e1}) and (\ref{eq:var1}) repeatedly to compute adjusted moments when making multiple observations. Clearly, if we were to adjust some parameters, $B$, based on information $D_{A,1}$ and $D_{A,2}$ and if we used (\ref{eq:e1}) and (\ref{eq:var1}) twice, first adjusting by $D_{A1,}$ followed by $D_{A,2}$ and then by $D_{A,2}$ followed by $D_{A,1}$, there is no reason to suppose that the final adjusted expectations and variances would be the same, and indeed in general they are not. In contrast, in equations (\ref{eq:ten}) and (\ref{eq:eleven}) the adjustments by $D_{A,1}$ and $D_{A,2}$ are summed and would, therefore, always result in the same adjusted expectations and variances, regardless of the order of updating.

Returning to our simple example, now suppose that we observe the survival times of patients in Group 1 one at a time. If we update the hazard rate for Group 1 sequentially for these patients using Bayes Theorem then the posterior distribution for the rate would not depend on the order in which we observed the survival time of the patients. However, if we propagated these changes one at a time using (\ref{eq:e1}) and (\ref{eq:var1}), then the final adjusted expectation and variance of the hazard rate in Group 2 would. In contrast, the updates would be commutative if (\ref{eq:ten}) and (\ref{eq:eleven}) were used to calculate the adjusted expectation and variance.

Other applications of Bayes linear kinematics are described by  \cite{Wils10} and \cite{Gos13}.

\section{Bayes linear kinematics in survival analysis}
\label{sec:blk}
\subsection{Introduction}

In Section \ref{sec:gam} we saw that the method of \cite{Gam91} combined Bayesian and Bayes linear updates in a dynamic survival model and used the structure of a Bayes linear Bayes model. This avoided the intensive numerical computations of a full Bayesian approach, but was not coherent in the sense that, if the order of the updating using the data was altered, the inferences would change. The same is true of dynamic generalised linear models \citep{West85}. In this section we outline our approach to this problem which retains the ease of calculation of the Gamerman approach by utilising a combination of full Bayesian and Bayes linear steps while solving the issue of commutativity by using Bayes linear kinematics to provide inferences which remain unchanged under permutations of the order of the observations.

In the following subsections we discuss in more detail the observational model and Bayesian updating, the specification and use of the guide relationship, the covariance structure defined by the system evolution and Bayes linear kinematic updating.  In particular, in Section \ref{guide}, we show how the conditions are satisfied for the existence of unique commutative Bayes linear kinematic updates.

\subsection{Observations, likelihood and Bayesian updating}
In each interval $R_{j}$ we give each of the $\lambda_{i,j}$ a gamma prior distribution which is conjugate to the conditional density and survival functions given in (\ref{density}) and (\ref{survival}). The distribution of the hazard $ \lambda_{i,j} $ for individual $ i $ in interval $ R_{j} $  is 
$ \lambda_{i,j} \sim\textrm{Ga}(\alpha_{i,j},\theta_{i,j}). $

The observations on different individuals $ i,\ i'$ are conditionally independent given $\{\lambda_{i,j}\}$ and $\{\lambda_{i',j}\}$. If individual $i$ is alive and uncensored at time $\tau_{j-1}$ then this individul can die in interval $R_{j}$, can survive interval $R_{j}$ or can be right-censored during interval $R_{j}$. The likelihood contribution from individual $i$ in interval $R_{j}$ is then
\begin{equation}
L_{i,j} = \lambda_{i,j}^{\delta_{i,j}}\exp\{-\lambda_{i,j}(t_{i,j}-\tau_{j-1})\}, \label{eq:likecontrib}
\end{equation}
where,
if individual $ i $ dies in $R_{j},$ we have $ \delta_{i,j}=1 $ and $ t_{i,j}=t_{i}$, 
if individual $i$ survives $R_{j},$ we have $ \delta_{i,j}=0 $ and $ t_{i,j}=\tau_{j},$
if individual $i$ is censored in $R_{j}$, we have $ \delta_{i,j}=0$ and $ t_{i,j}=t_{i}$ and
if $t_{i} < \tau_{j-1}$, then $ \delta_{i,j} =0 $ and $ t_{i,j}=\tau_{j-1}$.
Thus non-informative right censoring can be introduced into the model in a straightforward way. 

The complete likelihood contribution for individual $i$ is
\[ L_{i} = \prod_{j \in {\cal A}_{i}} L_{i,j}, \]
where $ {\cal A}_{i} = \{j\ :\ \tau_{j-1}<t_{i}\}$. This is proportional to the likelihood resulting from making observations $ D_{i,j}$ for $j=1, \ldots , J(i)$ where $ J(i) = \max(j\ :\ \tau_{j-1}<t_{i})$ if, given $ \{ \eta_{i,j} \}$, $ D_{i,1}, \ldots , D_{i,J(i)}$ were conditionally independent and, if $\delta_{i,j}=1$,  $D_{i,j}$ is an observation with value $ t_{i,j}-\tau_{j-1}$ from an exponential distribution with mean $ \lambda_{i,j}^{-1}$ and, if $\delta_{i,j}=0$, $D_{i,j}$ is a right-censored observation from the same distribution with censoring time $t_{i,j}-\tau_{j-1}$.  A Bayesian update with a fully-specified prior distribution and this equivalent likelihood could be done by conditioning on these exponential observations in any order. Since the likelihood is equivalent, this would also apply in our case, given a fully-specified prior distribution. This is convenient as it allows the posterior to be computed using the temporal structure of the dynamic model. Of course we wish the Bayes linear kinematic updates to have the same property and indeed they do.

The update for $\lambda_{i,j}$ is conjugate and so the posterior distribution for $ \lambda_{i,j}, $ based on the single observation $D_{i,j}$ at time $t_{i,j}$, is $\lambda_{i,j} \sim\textrm{Ga}(\alpha_{i,j,1},\theta_{i,j,1}), $
where
\[ \alpha_{i,j,1} = \alpha_{i,j,0}+\delta_{i,j}, \hspace{1cm} \theta_{i,j,1} = \theta_{i,j,0}+t_{i,j}-\tau_{j-1}. \]


In terms of the notation of Section \ref{sec:bl}, the elements of $ \mathcal{X}$ are $ \eta_{i,j}$, for $i=1, \ldots ,p$, and $ \beta_{j,k}$, for $k=0, \ldots ,q$, both for $j=1, \ldots ,J$. The information received, which causes us to change our beliefs about $ \eta_{i,j}$ is an observation of $ \mathcal{Z}_{i,j} =(\delta_{i,j},t_{i,j})$.
  
\subsection{Expectation and variance of $   \eta_{i,j}$}
\label{guide}
\cite{Gol04} used the mean and variance of a parameter indexing the conditional distribution of an observation $X$ to convey the information when $X$ is observed. For example, if $X \mid \lambda \sim {\rm Po}(\lambda)$, a Poisson distribution with mean $\lambda$, then the mean and variance of $\lambda$ could be used. However we require $ \lambda > 0$, we might typically expect a positive association between the means and variances of the various $ \lambda_{i,j}$ and an observation with $ \delta_{i,j}=1$ and $ t_{i,j}-\tau_{j-1}$ small can lead to an increase in the variance of $ \lambda_{i,j}$. These features make Bayes linear propagation of information between the different $ \lambda_{i,j}$ less appealing. To overcome them we can introduce a monotonic link function $g$ and use instead the mean and variance of $ \eta=g(\lambda)$, which, of course, also indexes the conditional distribution of $X$ since $ X \mid \lambda \sim {\rm Po}(g^{-1}(\eta))$. Like \cite{West85} and \cite{Gam91}, we propose $\eta = g(\lambda) = \log(\lambda)$, so that $ \eta $ is unbounded, the idea of the variance of $ \eta $ not depending on the mean seems more reasonable and, as we shall see, the variance of $ \eta $ can not be increased by an observation.

In fact, following \citep{West85}, we regard the link function as a {\em guide relationship}, written $\log(\lambda_{i,j}) \approx \eta_{i,j}$, which guides how information is passed between the individual hazards $ \lambda_{i,j} $ and the underlying parameters in the model. As part of our prior judgement we specify the conditional first and second moments of $ \{ \eta_{i,j} \}$, given the possible observations, using the guide relationship in conjunction with the conjugate update of the gamma distributions for $ \{ \lambda_{i,j} \}$. 

The expectation and variance of $\eta_{i,j}$ are then
\begin{equation}
\begin{array}{rcccc}
{\rm E}_{0}(\eta_{i,j}) & = & g_{1}(\alpha_{i,j},\theta_{i,j}) & = & f_{i,j}, \\ 
{\rm Var}_{0}(\eta_{i,j}) & = & g_{2}(\alpha_{i,j},\theta_{i,j}) & = & q_{i,j},
\end{array}\label{blkmv}
\end{equation}
for some functions $g_{1}(\cdot)$ and $g_{2}(\cdot)$, based on our guide relationship and providing a $ 1-1 $ transformation between $ (\alpha_{i,j}, \theta_{i,j}) $ and $ (f_{i,j}, q_{i,j}).$
In the prior specification we set $ \alpha_{i,j}=\alpha_{i,j,0} $ and $ \theta_{i,j} = \theta_{i,j,0}, $ giving $ f_{i,j}=f_{i,j,0} $ and $ q_{i,j}=q_{i,j,0}. $ 

This gives a posterior mean $ f_{i,j,1} $ and variance $ q_{i,j,1} $ for $\eta_{i,j}$, based on this single observation, $D_{i,j}$, using (\ref{blkmv}) but with the new parameter values. 

For Poisson observations, \cite{West85} suggest two possible choices for $ g_{1}$ and $g_{2}$. These are (i) use of the mean and variance of $\log\lambda_{i,j}$ and (ii) use of the mode and the curvature at the mode of the log density of $\log\lambda_{i,j}$.  A third possibility would be to equate the mean and variance of $ \lambda_{i,j}$ to the mean and variance of a lognormal distribution with parameters $ f_{i,j},q_{i,j}$. We will refer to these as the ``log-moment'', ``log-mode'' and ``lognormal'' methods respectively. In each case we find that
\begin{eqnarray*}
f_{i,j} = g_{1}(\alpha_{i,j},\theta_{i,j}) & = & h_{1}(\alpha_{i,j})-\log(\theta_{i,j}), \\ 
{\rm and} \hspace{1cm} q_{i,j} = g_{2}(\alpha_{i,j},\theta_{i,j}) & = & h_{2}(\alpha_{i,j})
\end{eqnarray*}
for some functions $ h_{1},\ h_{2}$. Observation of the survival, censoring or death of individual $i$ in interval $j$ gives the posterior expectation and variance
\[ f_{i,j,1}  =  h_{1}(\alpha_{i,j,0}+\delta_{i,j})-\log(\theta_{i,j,0}+t_{i,j}-\tau_{j-1}) \ \ {\rm and}\ \ q_{i,j,1}  =  h_{2}(\alpha_{i,j,0}+\delta_{i,j}). \]
Provided that $ h_{2}(z) $ is a strictly decreasing function of $ z $ for $ z>0$, $h_{2}(\alpha_{i,j,0}+1) < h_{2}(\alpha_{i,j,0})$ and the variance of $ \eta_{i,j} $ decreases when $ \delta_{i,j}=1 $ and remains the same when $ \delta_{i,j}=0$. This condition is satisfied by all three methods.

The inverse transformation is $ \alpha_{i,j}=h_{2}^{-1}(q_{i,j})$, where $ h_{2}^{-1}$ is the inverse function of $ h_{2}$, and $\theta_{i,j}=\exp(-f_{i,j})\exp\{h_{1}(\alpha_{i,j})\}$.

For the log-moment method, 
calculating the mean and variance of $\eta_{i,j}$ as those of $\log\lambda_{i,j}$ gives
\[ h_{1}(\alpha_{i,j})  =  \psi(\alpha_{i,j}) \ \  {\rm and}\  \
h_{2}(\alpha_{i,j})  =  \psi_{1}(\alpha_{i,j}), \]
where $\psi(\cdot)$ is the digamma function and $\psi_{1}(\cdot)$ is the trigamma function. We can solve $ q_{i,j}=\psi_{1}(\alpha_{i,j})$ numerically for $\alpha_{i,j}$ if required. 

For the log-mode method,
we set $ \mu_{i,j}=\log(\lambda_{i,j}) $ but, rather than using the mean and variance of $\mu_{i,j}$ directly, we say that $\eta_{i,j}$ is such that it has mean and variance given by
\[ f_{i,j}=m_{i,j},\ \ {\rm and}\ \ q_{i,j}=-\left\{\frac{d^{2}l_{i,j}(\mu_{i,j})}{d\mu_{i,j}^{2}}\right\}^{-1}_{m_{i,j}}, \]
where $m_{i,j}$ is the mode and $l_{i,j}(\mu_{i,j})$ is the log-density of $\mu_{i,j}$. We find
\[ h_{1}(\alpha_{i,j})  =  \log(\alpha_{i,j}) \ \  {\rm and}\  \
h_{2}(\alpha_{i,j})  =  \alpha_{i,j}^{-1}. \] 
Furthermore
\begin{displaymath}
\alpha_{i,j}=\frac{1}{q_{i,j}}\ \ {\rm and}\ \ \theta_{i,j}=\frac{1}{q_{i,j}}e^{-f_{i,j}}.
\end{displaymath}

For the lognormal method we set $ \alpha_{i,j}/\theta_{i,j}= \exp(f_{i,j}+q_{i,j}/2) $ and $ \alpha_{i,j}/\theta_{i,j}^{2}= \exp(2 f_{i,j} + q_{i,j})[\exp(q_{i,j}) -1] $ giving
\[ h_{1}(\alpha_{i,j})  =  \log\left[\alpha \sqrt{\alpha/(\alpha+1)} \right] \ \  {\rm and}\  \
h_{2}(\alpha_{i,j})  =  \log(1+\alpha^{-1}). \] 
Furthermore
\begin{displaymath}
\alpha_{i,j}=\left[\exp(q_{i,j})-1\right]^{-1}\ \ {\rm and}\ \ \theta_{i,j}=\left[\exp(q_{i,j})-1\right]^{-1}e^{-q_{i,j}/2}e^{-f_{i,j}}.
\end{displaymath}

\begin{figure}
\begin{center}

\resizebox{0.8\textwidth}{!}{\includegraphics{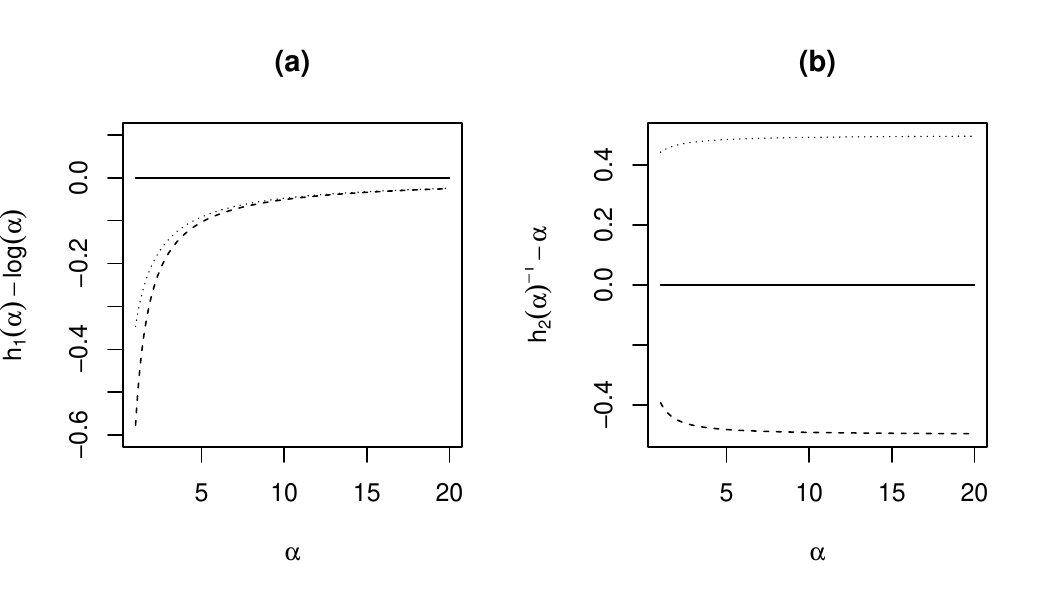}}

\end{center}

\caption{Differences between the functions (a) $ h_{1}(\alpha) $ and (b) $ h_{2}(\alpha)^{-1} $ ({\it ie} the precision of $ \eta$) plotted against $\alpha^{(0)}$ for the log-moment method (dashes), the log-mode method (solid) and the lognormal method (dots). The value for the log-mode method is subtracted as a reference.}
\label{fig:compare1}
\end{figure}

While we are free to choose any of these methods and we will use the log-mode method in the example in Section \ref{sec:ex}, a brief comparison is useful to investigate the sensitivity of the results to this choice. Figure \ref{fig:compare1} (a) shows the differences between $ h_{1}(\alpha)$ for the three methods with the log-mode method as a reference. We see that the curves for the log-moment and lognormal methods are very similar, lying slightly below the curve for the log-mode method and converging towards it as $ \alpha $ increases. Figure \ref{fig:compare1} (b) shows the differences in the reciprocal of $ h_{2}(\alpha)$, that is the precision of $ \eta$, again with the log-mode method, where $ h_{2}(\alpha)^{-1}=\alpha$, as a reference.  The precision with this method would increase by  1 when a death is observed. The precision using the log-mode method lies between those for the other two methods with the log-moment and lognormal methods giving respectively smaller and larger values. The differences between adjacent curves are slightly less than 0.5.

As a second comparison, let us return to our simple example. We have two hazards, $\lambda_{1},\lambda_{2}$. For $ i=1,2$, $ \lambda_{i}$ has a gamma Ga$(\alpha_{i}^{(0)},\ \theta_{i}^{(0)})$ distribution. We give $ \eta_{1}$ and $ \eta_{2}$ a correlation $\rho$. We make a single observation with hazard $ \lambda_{1}$ and compare the resulting adjusted mean and standard deviation for $\lambda_{2}$. We make this comparison for (i) an observed death and (ii) a censored observation, both at time $ t_{1}-\tau=t$ since the beginning of the interval. 
The adjusted distribution of $ \lambda_{i}$ is Ga$(\alpha_{i}^{(1)},\ \theta_{i}^{(1)})$. The prior mean and variance of $ \lambda_{i}$ are $ m_{i}^{(0)} = \alpha_{i}^{(0)}/\theta_{i}^{(0)} $ and $ v_{i}^{(0)} = \alpha_{i}^{(0)}/(\theta_{i}^{(0)})^{2}$ respectively and the corresponding adjusted values are $ m_{i}^{(1)} $ and $ v_{i}^{(1)}$. For the censored observation, $ \alpha_{1}^{(1)}=\alpha_{1}^{(0)} $ and $ \theta_{1}^{(1)}=\theta_{1}^{(0)}+t$ and it is simple to show that, for all three of our Bayes linear kinematic methods, $  \alpha_{2}^{(1)}=\alpha_{2}^{(0)} $ and $ \theta_{2}^{(1)}=\theta_{2}^{(0)}k_{c}$ where $ k_{c} = \{ \theta_{1}^{(1)}/ \theta_{1}^{(0)} \}^{\rho}$ and therefore $ m_{2}^{(1)}=m_{2}^{(0)}/k_{c}$ and $ v_{2}^{(1)}=v_{2}^{(0)}/k_{c}^{2}$. For the observed death, $ \alpha_{1}^{(1)}=\alpha_{1}^{(0)}+1$ and the results depend on the choice of transformation.

\begin{figure}
\begin{center}

\resizebox{\textwidth}{!}{\includegraphics{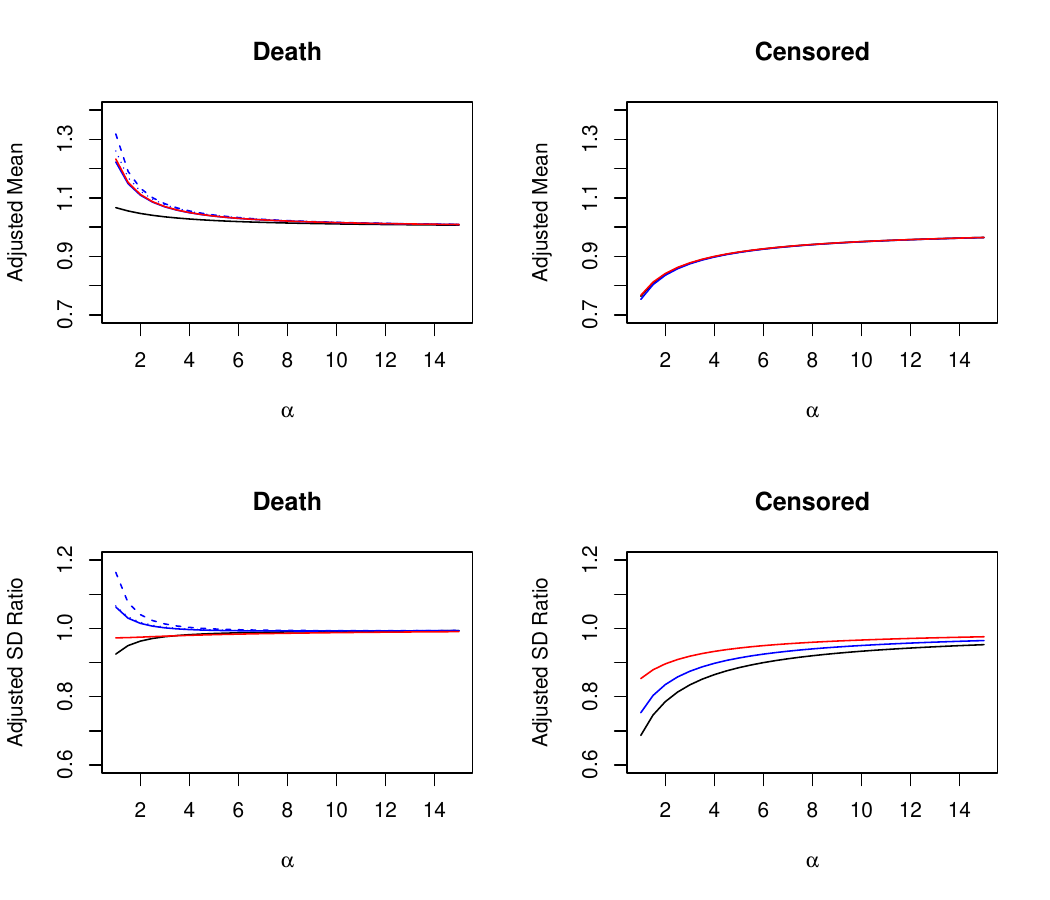}}

\end{center}

\caption{Adjusted means and scaled standard deviations of $ \lambda_{2}$ given a single observation with hazard $ \lambda_{1}$. Results for the three Bayes linear kinematic methods with log link are shown in blue with dashes for the log-moment method, solid for the log-mode method and dots for the lognormal method. The black line is for a full-Bayes posterior and the red line is for Bayes linear kinematics with identity link.}
\label{fig:compare2}
\end{figure}

Figure \ref{fig:compare2} shows an example where we set $ \alpha_{1}^{(0)}=\alpha_{2}^{(0)} $ and $ \theta_{i}^{(0)}=\alpha_{i}^{(0)}$ and the correlation between $ \eta_{1}^{(0)} $ and $ \eta_{2}^{(0)} $ is 0.7. The prior mean of $ \lambda_{i}$ is 1 for all $ \alpha_{i}^{(0)} $ but the prior standard deviation depends on $ \alpha_{i}^{(0)}$ so the adjusted standard deviations have been scaled by dividing by the prior standard deviations. For comparison, the posterior mean and standard deviation resulting from a full-Bayes analysis are shown. A bivariate normal prior was used for $ \eta_{1}$ and $\eta_{2}$, giving $ \lambda_{i}$ a lognormal prior but the parameters were chosen to match the mean and variance of the prior gamma distribution for $ \lambda_{i}$ in the Bayes linear kinematic case. Also shown is the result of a Bayes linear kinematic update directly based on $ \lambda_{i} $ rather than {\it via} $ \eta_{i}$, with a correlation of 0.7 between $ \lambda_{1}$ and $ \lambda_{2}$, which we term the ``identity-link'' method. In every case we set $ t=1/\lambda^{*}$ where $ \lambda^{*}$ is one prior standard deviation greater than the prior mean of $ \lambda_{i}$.

In the case of an observed death, we see that the Bayes linear kinematic methods all give very similar adjusted means which are slightly greater than the full-Bayes posterior mean for small $ \alpha=\alpha^{(0)}_1=\alpha^{(0)}_2$ but quickly converge towards it as $ \alpha $ increases. As $\alpha$ increases the shape of a gamma distribution becomes more similar to that of a lognormal distribution. Of these three curves, log-mode is the lowest and log-moment the highest.  The adjusted standard deviations give a similar picture except that the identity link method gives results closest to full-Bayes.

In the case of a censored observation, the three Bayes linear kinematics (BLK) with log link methods give identical results. In fact, in the adjusted mean, the other two methods give results which are almost identical to these three. The means for the identity link and log link BLK methods are respectively slightly greater and slightly less than the full-Bayes means. There is more of a difference in the standard deviations with the log link BLK method being closer to full Bayes than the identity link BLK method.

We prefer to use a log link method rather than an identity link because of the guarantee of positive $ \lambda$, the non-increasing variance and the variance stabilisation. Of the three log link BLK methods, the log-mode method is the simplest, comes closer to full Bayes than the log-moment method and, while its results are very similar to those of the lognormal method, they are very slightly closer to full Bayes. On these grounds, the log-mode method is our preferred method.

\subsection{System evolution}
We relate the log hazards $ \eta_{i,j} $ to model parameters $\vect{\beta}_{j}\ (j=1, \ldots ,r)$ using $ \eta_{i,j}=    \vect{x}_{i}'   \vect{\beta}_{j} $
where $     x_{i}' = (1, x_{i,1}, \ldots , x_{i,q}),$  $     \beta_{j}' = (\beta_{j,0}, \ldots \beta_{j,q}) $ and $ \beta_{j,0}= \log \lambda_{0,j}. $

To build the prior specification for the parameter vector,  \cite{Gam91} utilised the system evolution of a  dynamic generalised linear model \citep{West85}. Here we extend this approach and incorporate some ideas from \cite{Far03} to give a more general form.

We write
\[  \beta_{j}-    B_{j}=G_{j}(   \beta_{j-1}-     B_{j-1})+   \epsilon_{j}, \]
where the vector $   \epsilon_{j}$ is the cumulative innovation over $R_{j}$ which has mean zero and covariance matrix $E_{j}$. \cite{Gam91} suggests allowing both the system evolution matrix $ G_{j}$ and the innovation covariance matrix $E_{j}$ to depend on the length $b_{j}$ of the interval $R_{j}$, with $E_{j}=b_{j}\bar{E}_{j}$, where $\bar{E}_{j}$ is a ``unit'' covariance matrix. We introduce $ B_{j}=m_{j}+U$, where $ m_{j}$ is a specified mean vector and the vector $U$, with mean zero and covariance matrix $ C_{0}$, allows us to specify a global component in our uncertainty about $ \beta_{1}, \ldots , \beta_{r}$.

 Beliefs about the parameter vector $   \beta_{j}$ are not given a full distributional form but are simply specified in terms of a mean vector $     m_{j} $  and a covariance matrix $ C_{j}.$ We write $    \beta_{j}\sim[    m_{j},C_{j}].$ Thus if our prior beliefs for the parameter vector in interval $ R_{1}$, $   \beta_{1}=(\beta_{1,0},\ldots,\beta_{1,q})$, are given by $    \beta_{1}\sim[    m_{1},C_{1}], $ then we can calculate the prior specification for the parameters in interval $R_{j}$ as $    \beta_{j}\sim[    m_{j},C_{j}], $ where the mean vector is $ m_{j}$ and the covariance matrix is
\[ C_{j} = C_{0} + G_{j}(C_{j-1}-C_{0})G_{j}^{'}+E_{j}, \]
as $   \beta_{j-1}$ and $   \epsilon_{j}$ are independent. The covariance matrix between parameter vectors in different intervals is given by
\begin{displaymath} 
{\rm Cov}(   \beta_{j},   \beta_{j+l}) = C_{0}+(C_{j}-C_{0})\prod_{m=j+1}^{j+l}G_{m}=C_{j,j+l}. 
\end{displaymath}
For example the covariance matrix between $   \beta_{j-1}$ and $   \beta_{j}$ is ${\rm Cov}(   \beta_{j-1},   \beta_{j}) =  C_{0} + (C_{j-1}-C_{0})G_{j}$. 

Using the guide relationship, the prior expectation and variance of $\eta_{i,j}$ are
\begin{displaymath}
\begin{array}{cclcc}
f_{i,j} & = & {\rm E}_{0}[\eta_{i,j}] & = &     x_{i}'    m_{j}, \\
q_{i,j} & = & {\rm Var}_{0}(\eta_{i,j}) & = &     x_{i}'C_{j}    x_{i}.
\end{array}
\end{displaymath}
The prior covariances between the transformed quantities $ \eta_{i,j}$ are given by
\[ {\rm Cov}_{0}(\eta_{i,j},\eta_{k,l})  =      x_{i}'C_{j,l}    x_{k} = q_{(i,k)(j,l)} \]
where $ C_{j,j}=C_{j} $ and $ q_{(i,i)(j,j)} = q_{i,j}. $
Finally we need the covariances between the transformed quantities and the parameter values. These are
\[   {\rm Cov}_{0}(\eta_{i,j},   \beta_{l})  =      x_{i}'C_{j,l} =     s_{i(j,l)}. \]

 We define $    \Omega=(   \eta',   \beta')^{'}$, where $   \eta=(\eta_{1,1}, \ldots, \eta_{p,r})^{'}$ and  $   \beta=(\beta_{1,0}, \ldots, \beta_{r,q})^{'}$ to be the set of all quantities of interest. Prior specifications for this set are given by ${\rm E}_{0}(\Omega)=    L$ and
\begin{displaymath}
{\rm Var}_{0}(\Omega)=\left(
\begin{array}{cc}
{\rm Var}_{0}(   \eta) & {\rm Cov}_{0}(   \eta,   \beta) \\
{\rm Cov}_{0}(   \beta,   \eta) & {\rm Var}_{0}(   \beta)
\end{array}
\right)=Z,
\end{displaymath}
and each of the components of the matrix can be calculated in terms of the quantities found above.

\subsection{Bayes linear kinematics}

Having made observations, we propagate changes in belief about individuals through to the other individuals, other intervals and model parameters, through the Bayes linear structure, using Bayes linear kinematics. 
Adjusting our beliefs about $ \Omega $ having observed individual $ i $ in interval $ R_{j} $ gives an adjusted expectation and variance for $ \Omega$ of
\begin{displaymath}
{\rm E}_{1(i,j)}(    \Omega)=    L+{\rm Cov}_{0}(    \Omega,\eta_{i,j})\frac{[f_{i,j,1}-f_{i,j,0}]}{q_{i,j,0}},
\end{displaymath}
and
\[ {\rm Var}_{1(i,j)}(    \Omega)  =  Z-{\rm Cov}_{0}(    \Omega,\eta_{i,j}){\rm Cov}_{0}(\eta_{i,j},    \Omega) \left(\frac{1}{q_{i,j,0}}-\frac{q_{i,j,1}}{q_{i,j,0}^{2}} \right). \]
As discussed in Section \ref{sec:bl}, we need to define a Bayes linear kinematic update for multiple observations. If we simply performed the updates given in the equations above sequentially, for each patient in each interval,  as in \cite{Gam91}, then the final values obtained for the adjusted moments of the unknown quantities would change depending on the order in which the updates are performed. The order of patients within an interval should be irrelevant and, in fact, since the analysis is retrospective in contrast to a sequential forecasting algorithm in time-series analysis, the order in which intervals are entered should be irrelevant.
When a unique commutative Bayes linear kinematic update exists the adjusted expectation and variance are
\begin{displaymath}
{\rm Var}_{p^{*}}(    \Omega)=\left\{\sum_{j=1}^{r}\sum_{i=1}^{p_{j}}{\rm Var}_{1(i,j)}^{-1}(    \Omega)-(p^{*}-1){\rm Var}_{0}^{-1}(    \Omega)\right\}^{-1}
\end{displaymath}
and
\[ {\rm E}_{p^{*}}(    \Omega)  =  {\rm Var}_{p^{*}}(    \Omega)\left\{\sum_{j=1}^{r}\sum_{i=1}^{p_{j}}{\rm Var}_{1(i,j)}^{-1}(    \Omega){\rm E}_{1(i,j)}(    \Omega)-(p^{*}-1){\rm Var}_{0}^{-1}(  \Omega){\rm E}_{0}(    \Omega)\right\}, \]
where $p^{*}=\sum_{j=1}^{r}p_{j}$ and $ p_{j} $ is the number of patients known to be alive at time $ \tau_{j-1}.$ From these we can obtain the adjusted means and variances of the parameters. 

With any of the log link BLK methods, $ q_{i,j}$, the variance of $ \eta_{i,j}$, decreases if individual $ i $ dies in interval $R_{j}$ and stays unchanged by any other observation on individual $i$ in $R_{j}$. We can therefore use the commutative formulae (\ref{eq:ten}) and (\ref{eq:eleven}). Furthermore, propagation of a decrease in the variance of $\eta_{i,j}$ through the Bayes linear structure can not increase, but can decrease, the variance of other $\eta_{i',j'}$. It follows that, provided that we observe at least one death, however we partition the information received, at least one contribution causes a decrease in at least some of the variances and no contribution can cause an increase in any variance. The update given by (\ref{eq:ten}) and (\ref{eq:eleven}) is therefore the unique commutative update. See \cite{Gol04} and \cite{Wils10}.

\subsection{Prior Robustness}
There remain questions around robustness with respect to prior parameters. We now investigate this in the context of our simple example, previously described, in which we are interested in the survival times of two groups of patients, each with a constant hazard rate and a gamma prior distribution. The prior parameters in this example are $(\alpha_1^{(0)},\theta_1^{(0)},\alpha_2^{(0)},\theta_2^{(0)},\rho)$. To investigate robustness with respect to the prior specifications we perform a simulation exercise in which we simulate observations $t_{1,j}\sim\textrm{Exp}(\lambda_1)$, $j=1,\ldots,N$ with $\lambda_1=\alpha/\theta$ for chosen $(\alpha,\theta)$. We assume that the prior parameters are of the form $\alpha_1^{(0)}=\alpha_2^{(0)}=k\alpha$ and $\theta_1^{(0)}=\theta_2^{(0)}=\sqrt{k}\theta$, so that $k$ changes the prior means but not the prior variances of $\lambda_1,\lambda_2$. We vary $k$ to observe the effect of changes in the prior means of Groups 1 and 2. The observations are made on Group 1 and we consider the effect on the posterior mean of Group 2. We consider different numbers of observed survival times ($N$) between 2 and 1000. For each of these sample sizes the simulation is run 1000 times.

\begin{figure}
\centering
\resizebox{0.6\textwidth}{!}{\includegraphics{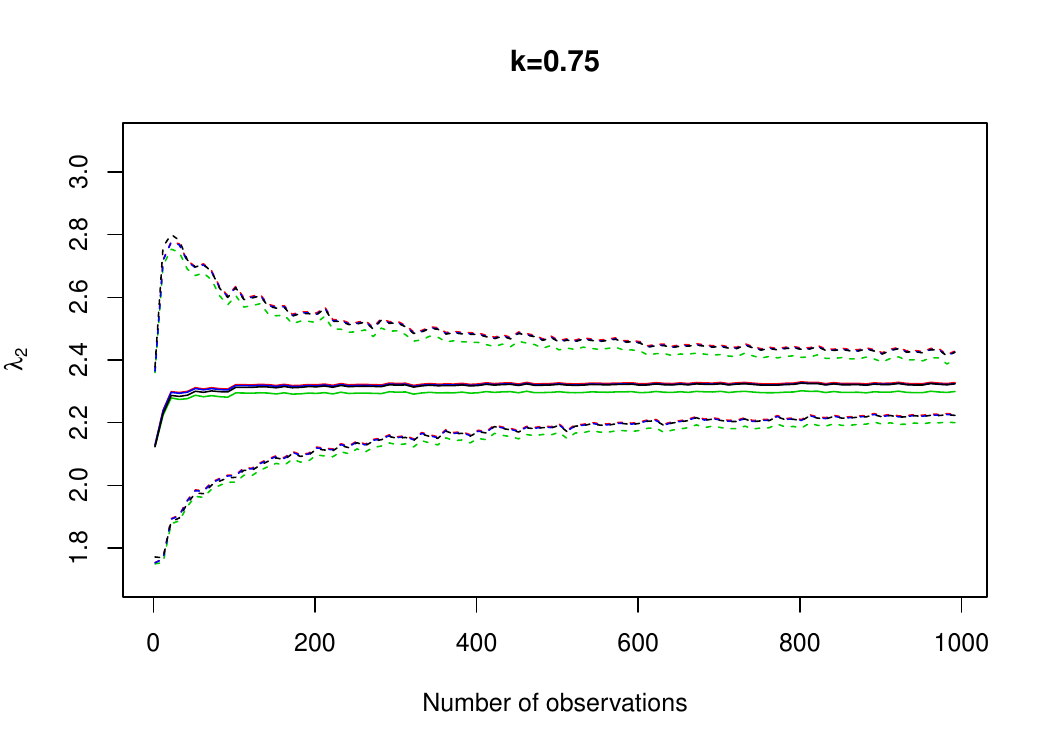}}
\resizebox{0.6\textwidth}{!}{\includegraphics{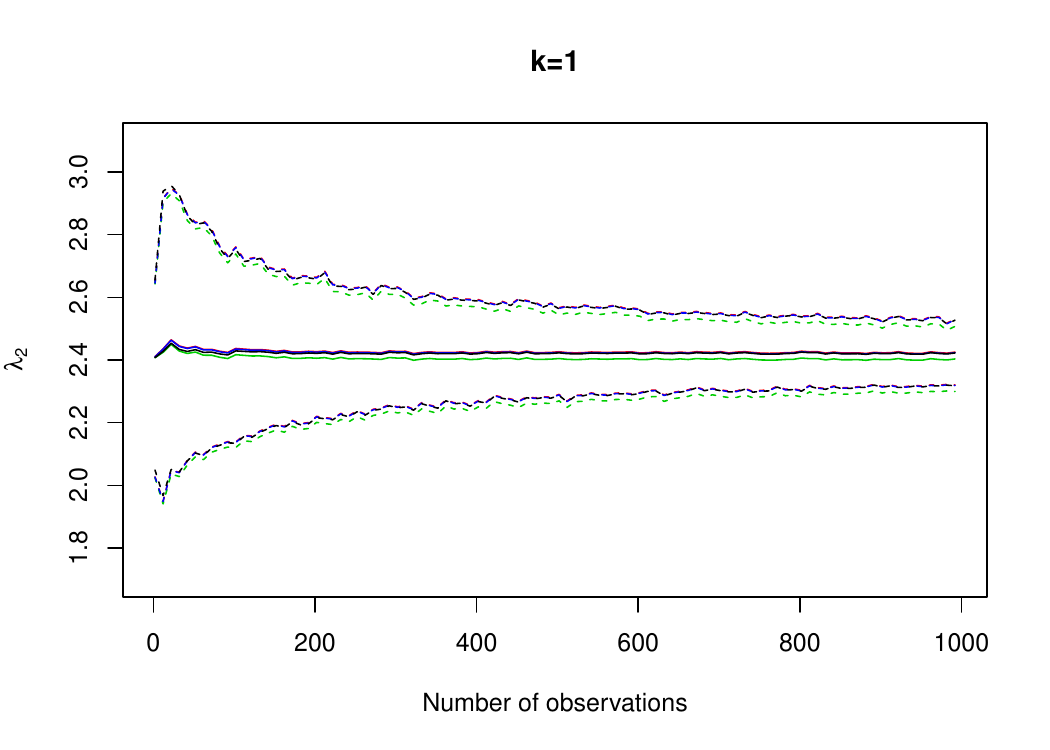}}
\resizebox{0.6\textwidth}{!}{\includegraphics{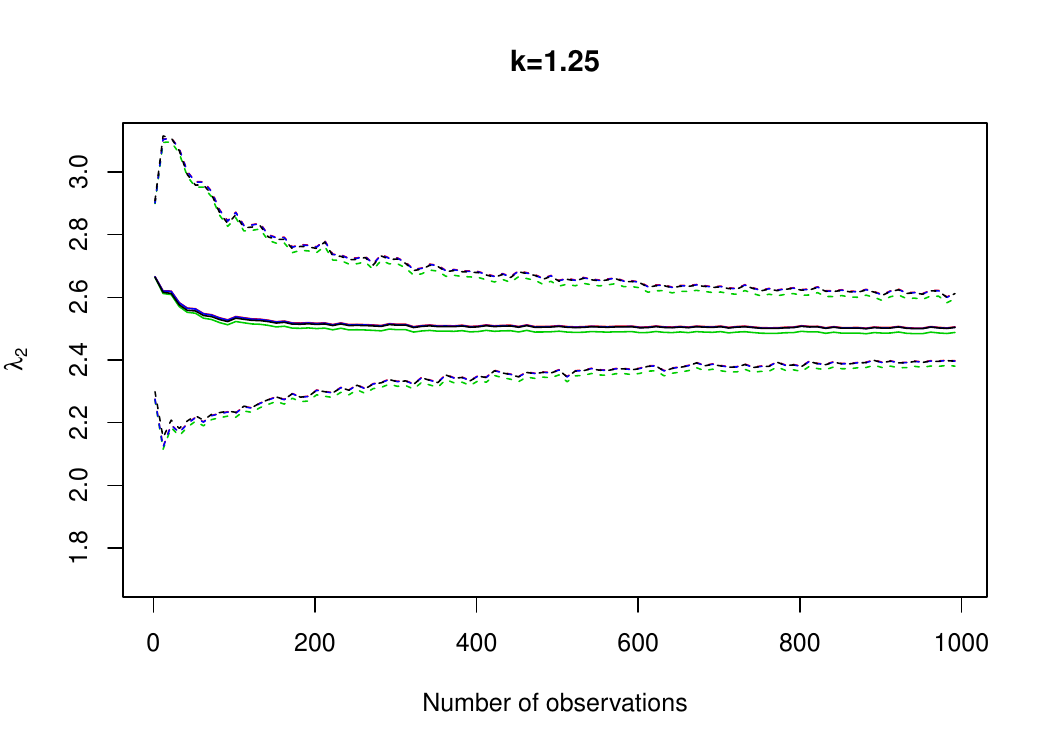}}

\caption{The sample size plotted against the empirical mean over 1000 simulations of the posterior mean of $\lambda_2$. Also given are 95\% empirical bounds for the posterior mean from the 1000 runs (dashed lines). We compare three transformations; the log-moment (red), log-mode (green) and lognormal (blue) methods and also a full-Bayes analysis (black).}
\label{Prior1}
\end{figure}

We choose $\alpha=12, \theta=5$ and $\rho=0.7$. In Figure \ref{Prior1}, we see the sample size plotted against the empirical mean over the 1000 runs of the posterior mean of $\lambda_2$. Also given are 95\% empirical bounds for the posterior mean from the 1000 runs. We compare the three transformations identified earlier; the log-moment (red), log-mode (green) and lognormal (blue) methods. In addition we show the results from a full-Bayes analysis, using numerical integration, with a lognormal model in black. The plots, from the top, show the cases $k=0.75$, $k=1$ and $k=1.25$.

Between the three plots we see a pattern. The log-moment, lognormal and full-Bayes methods give results which are almost identical. However the log-mode method results in posterior means which are lower than those of the other methods.

The simulated value for $\lambda_2$ is 2.4. When $k=1$, all  methods give posterior means which are close to this value for sample size above 100. Even with smaller samples, however, the posterior mean is on average close to 2.4. The variation in the posterior means decreases as sample size increases.  When $k=1.25$, the prior mean is higher than the hazard rate and all of the methods tend to give posterior means greater than 2.4 for all sample sizes but they all give posterior means closer to 2.4 as the sample size increases. The converse is true for $k=0.75$, with all of the methods tending to give posterior means smaller than 2.4. Note, however, the closeness to the results from the full Bayesian analysis.

\begin{figure}
\centering
\resizebox{\textwidth}{!}{\includegraphics{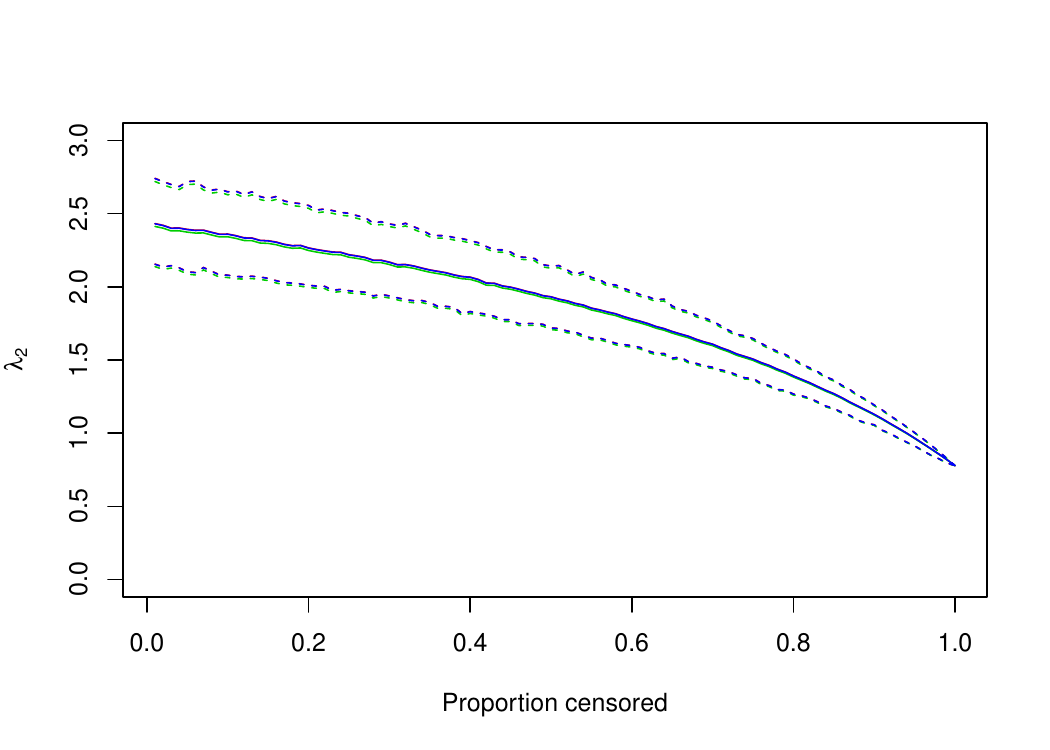}}
\caption{The proportion of right censored observations plotted against the empirical mean over the posterior means of $\lambda_2$ from 1000 samples. Also given are the 2.5\% and 97.5\% quantiles for the posterior mean from the 1000 runs (dashed lines). We compare three transformations; the log-moment (red), log-mode (green) and lognormal (blue) methods.}
\label{Cens}
\end{figure}

We can also investigate the impact of censoring on the posterior mean for this example. In this case, we use the same prior as for Figure \ref{Prior1} but hold the number of observations constant at 100 and vary the proportion of those observations which are right censored. All of the right censoring occurs at $t=0.2$ which is a little less than half of the predictive mean lifetime of $2.4^{-1} =0.417$. Each simulated data set was formed by creating the required number, say $m$, of observations censored at $t=0.2$ and then simulating $100-m$ uncensored observations.  The fraction of censored observations was therefore fixed at $m/100$. In Figure \ref{Cens} we compare the three transformation methods and plot the average  and 95\% symmetric intervals for the posterior means from 1000 samples. On the x-axis is the proportion of right censored values.

The posterior means of $\lambda_{2}$ decrease as the censoring proportion increases because the posterior means of $\lambda_{1}$ decrease. This is because the number of observed deaths decreases and the proportion of cases surviving beyond $t=0.2$ increases. We see that there are small differences between the log-mode method and the other two when there is very little censoring in the data but as censoring increases this method converges to the other two. We also see less variation in the posterior means across the runs for all methods as the proportion of censoring increases. This is because the variation in the data decreases.

\section{Application: Leukemia survival rates}
\label{sec:ex}
\subsection{Background}

\cite{Hend02} investigated leukemia survival rates for patients based on data obtained from the North-West Leukemia Register in the UK. In their analysis they considered adult patients with acute myeloid leukemia, the most common type, of which there are 1043 cases in the database between 1982 and 1998. Of these 1043 patients, 879 died and 164 were right censored. The response variable in the study is the time $T$ in days until the death of a patient.

The database also includes the values, for each patient, of a number of covariates which are thought to affect the survival rates of patients with leukemia. As this is an observational study, inferred effects are interpreted as predictive and not necessarily causal. The covariates are
\begin{enumerate}
 \item The age $A_{i}$ in years of patient $i$. We use $x_{i,1}=A_{i}-60$ so that the intercept refers to a typical patient.
 \item The sex of the patient. That is, $x_{i,2}=-1$ if the patient is female and $x_{i,2}=1$ if the patient is male.
 \item White blood cell count (WBC) at the time of diagnosis. This is truncated at 500 units with 1 unit$=50\times10^{9}/$L. We use $x_{i,3}=WBC-8$.
 \item The Townsend score, a measure of the deprivation of an area of residence on a scale from -7 to 10. The larger this is, the less affluent the area \citep{Town88}. The Townsend score is used directly as the covariate $x_{i,4}$. 
\end{enumerate}
The hazard for patient $i$ in interval $R_{j}$ is then $\exp\{\beta_{j,0}+\sum_{k=1}^{4}\beta_{j,k}x_{i,k}\}$. The hazards $\exp\{ \beta_{j,0}+\beta_{j,2}\}$ and $\exp\{ \beta_{j,0}-\beta_{j,2}\}$ refer respectively to a male patient and a female patient each aged 60, with a white blood cell count of 8 units and a deprivation score of 0.

\subsection{Elicitation of prior information}
We need to choose the time intervals to be used. In a case like this, with a relatively small proportion of censoring so that most of the recorded times are deaths, the times become much less frequent later. We might suppose {\it a priori} that the times will have roughly an exponential distribution.  Thus it does not seem appropriate to use equal time intervals. Instead, suppose we define
$ \tau_{j}=-\nu\log(1-\kappa j),$
for some values of $\nu$ and $\kappa$. We choose $\kappa=0.1$ to give ten intervals and $\nu=500$, an approximate prior expectation of the marginal mean lifetime, with the aim of having approximately $10\%$ of the events in each interval. 

Specification of prior beliefs is required in the form of prior means, variances and covariances for the collection of parameter vectors $(   \beta_{1}, \ldots ,   \beta_{r})$ where $\beta_{j}=(\beta_{j,0},\beta_{j,1},\beta_{j,2},$ $\beta_{j,3},$ $\beta_{j,4})$. Within our formulation this is done by specifying the moments for $   \beta_{1}$, the system evolution matrices  $G_{j}(b_{j})$ and the covariance matrices, $E_{j}$, for the innovations $\epsilon_{j}$.

We can obtain suitable prior moments for $ \beta_{1,0} $ by assuming a constant hazard and considering a plausible range for the mean lifetime for ``baseline'' patients. A mean of -6 and a standard deviation of 0.8 for $ \beta_{1,0}$ gives a $\pm 2$ standard deviations range corresponding to a range for the mean lifetime from 81 to 1998 days.
In order to specify values for the moments of the coefficients of the covariates we utilise the proportional hazards assumption. If hypothetical individuals $i$ and $i'$ have covariate vectors which are equal except that $ x_{i,k} \neq x_{i',k} $ and their hazards are $h_{i}(t)$ and $h_{i'}(t)$ respectively for $ \tau_{j-1}\leq t < \tau_{j},$ then their ratio is
\begin{displaymath}
r_{i,i'}(t)=\frac{h_{i}(t)}{h_{i'}(t)}=\exp\{\beta_{j,k}(x_{i,k}-x_{i',k})\}.
\end{displaymath}
That is, we can specify values for the hyperparameters by eliciting information about ratios of hazards between individuals. This can be done by considering the situation where both individuals are alive at time $ \tau_{j-1} $ and we are told that one of them dies during the interval. The conditional probability that it is individual $ i $ who dies first is then $ L_{C,i,i'}(t)=h_{i}(t)/\{h_{i}(t)+h_{i'}(t)\}$, the Cox partial likelihood given just these two individuals \citep{Cox72} and $ r_{i,i'}(t)= L_{C,i,i'}(t)/\{1+L_{C,i,i'}(t)\}$ so quantiles for $ L_{C,i,i'}(t)$ are easily converted to quantiles for $r_{i,i'}(t)$ and for $\beta_{j,k}.$ 

For example,  consider $\beta_{1,1}$, the coefficient of the age of the patient. Ages in the investigation range from 14 to 92 and we might expect that increasing age would increase the hazard. We elicit the mean and variance of $\beta_{1,1}$ by supposing that patient $i$ is 10 years older than patient $i'$. Then the ratio of hazard functions for the two patients is
\begin{displaymath}
\frac{h_{i}(t)}{h_{i'}(t)}=\exp\{10\beta_{1,1}\},
\end{displaymath}
as long as individuals $i$ and $i'$ have identical covariates otherwise. A $20\%$ decrease in the hazard suggests $\beta_{1,1} \approx  0.005$ and an $80\%$ increase in the hazard suggests $\beta_{1,1} \approx 0.06$. If we assume that these two values give an approximate 95\% interval for a normal prior distribution over $\beta_{1,1}$, then we obtain
\begin{displaymath}
{\rm E}_{0}(\beta_{1,1})=0.02,~~{\rm Var}_{0}(\beta_{1,1})=0.0004.
\end{displaymath}
We can perform a similar process for $\beta_{1,2},\beta_{1,3}$ and $\beta_{1,4}$. Details are omitted.  The results of the elicitation process are given in Table \ref{tab:KJW}.
\begin{table}
\caption{Prior moments for each of the effects.}
\label{tab:KJW}

\begin{center}
\begin{tabular}{lcrll}
\hline
Effect & & Mean & Std.Dev. & Variance \\ \hline
Baseline & $\beta_{1,0}$ & -6.000 & 0.80 & 0.64 \\
Age  & $ \beta_{1,1} $ & 0.020 & 0.02 & 0.0004 \\
Sex & $ \beta_{1,2} $ & 0.000 & 0.35 & 0.1225 \\
WBC & $ \beta_{1,3} $ & 0.005 & 0.005 & 0.000025 \\
Deprivation score & $ \beta_{1,4} $ & 0.000 & 0.1 & 0.01 \\
\hline
\end{tabular}
\end{center}
\end{table}

\cite{Rev10} discuss the problem of assessing prior covariances in Bayes linear structures. In our case, we centred the covariates by subtracting typical values so that the baseline hazard refers to a central, ``typical'', case. Having done this, it is reasonable to have zero prior covariance between the coefficients of covariates in an interval and the baseline log hazard in that interval. Moreover, we saw no reason to suppose that learning that the coefficient of one covariate was larger than our prior expectation of it would cause us to revise our expectation for the coefficient of another covariate so we set the prior covariances between coefficients within a time interval to zero. It still remains to make specifications of the forms of $G_{j}$ and $E_{j}$. 
With equal-length intervals, a prior in which $ E_{j} $ decreases with $ j $ is likely to be appropriate as we may well expect effect sizes to settle towards steady values as time increases. However, with our lengthening intervals, we propose a stationary prior. We set $     B_{j} =     B$ with $\E(B) =     m_{0},\ \Var(B)=C_{0},\ E_{j}=E $ and $ G_{j} = \phi I_{5} $ for all $ j, $ where $\phi$ is a scalar with $0<\phi<1$ and $I_{5}$ is the $(5\times 5)$ identity matrix. 
We achieve stationarity by setting $C_{j}=C=C_{0}+E/(1-\phi^{2})$. Once we have chosen values for $C,\ C_{0}$ and $\phi$, this allows us to determine the value of $E$.

The choices of $C_{0},\ E$ and $ \phi $ govern the prior covariances between coefficient values in different intervals and hence the degree of smoothing over time. Let $\Gamma_{k}$ be the symmetric covariance matrix between $ \beta_{j}$ and $\beta_{j+k}$. Then $\Gamma_{k}=C_{0}+\phi^{k}(C-C_{0})$. In particular, in the special case where $C_{0}$ and $E$ are diagonal and $C_{0}=c_{0}C$ for some scalar $c_{0}\in [0,1]$, then $\Gamma_{k}=\phi_{k}C$ where $ \phi_{k}=c_{0}(1-\phi^{2})+\phi^{k}$. If we were to learn the value of $ \beta_{j}$ then our covariance matrix for $\beta_{j+k}$ would be reduced from $C$ to $C-\Gamma_{k}C^{-1}\Gamma_{k}$ or $C(1-\phi_{k}^{2})$ in the special case. By considering such reductions in variance, suitable values may be elicited. Note that it may be preferred to consider, rather than complete learning of the value of $ \beta_{j}$, the effect of gaining information which reduces the variance of $ \beta_{j}$ from $C$ to $(1-\delta)C$ for some $\delta \in [0,1]$. Then, using (\ref{eq:var1}), the covariance matrix of $ \beta_{j+k}$ becomes $ C-\delta \Gamma_{k} C^{-1} \Gamma_{k} $ or, in the special case, $C(1-\delta \phi_{k}^{2})$. In this example we choose the special case with $ c_{0}=0$ and $ \phi=0.92 $ to give moderately strong correlation between neighbouring intervals, falling to quite weak correlation between the first and last. Specifically, this gives $\phi_{1}^{2}=0.846,\ \phi_{5}^{2}=0.434$ and $ \phi_{9}^{2} = 0.223$. 

\subsection{Results}
For the reasons given in Section \ref{guide}, we shall use the log-mode method here for all calculations.

Having updated the four parameters of interest using the data we can plot the effects of the covariates over time. The posterior means for the effects of age and sex are shown by circles in Figure \ref{Bayes}. The posterior parameter means for interval $ j $ are plotted at $ \tau^{*}_{j} = -\nu \log\{1-\kappa(j-0.5)\} $ but would remain constant within that interval. The horizontal scale is proportional to $ 1- \exp(-t/\nu)$. Also plotted are $\pm2$ standard deviation intervals.






The effect of age is marked. There is also a strong suggestion of a  temporal component to the effect of age. The posterior means for the coefficients of age are positive in all intervals, as are most of those for sex, indicating that increased age and being male both increase the hazard associated with death from acute myeloid leukemia.

The age effect is strongest initially and appears to decrease over time then finally increase again. Only one of the $\pm2$ standard deviation intervals includes zero, however, so even in intervals where the effect of age is not as strong, there is still an effect. In contrast, the effect associated with sex appears to increase over time. The effect of this covariate is less clear, however, as all but two of the $\pm2$ standard deviation intervals contain zero.

The posterior means for the effects of all covariates in all intervals are given in Table \ref{postleuk}. Posterior standard deviations are given in brackets.

\begin{table}
\caption{Posterior means and standard deviations for each of the parameters in each interval.}
\label{postleuk}

\begin{center}
\begin{tabular}{rrrrrr}\hline
$j$ & $ \tau_{j} $ & $\beta_{j,1} \times 10^{2}$ & $\beta_{j,2} \times 10$ & $\beta_{j,3} \times 10^{3}$ & $\beta_{j,4} \times 10^{2}$ \\ \hline
 1 &   52.6 & 3.647 (0.401) & -0.047 (0.554) &  6.453 (0.584) & 3.492 (1.557)\\
 2 &  111.6 & 3.743 (0.454) &  0.694 (0.678) &  3.541 (0.975) & 5.472 (1.891)\\
 3 &  178.3 & 2.689 (0.468) &  0.370 (0.833) &  3.040 (1.136) & 2.172 (2.493)\\
 4 &  255.4 & 2.424 (0.480) &  1.170 (0.861) &  4.086 (1.104) & 3.940 (2.339)\\
 5 &  346.6 & 2.126 (0.505) &  0.749 (0.896) &  3.272 (1.421) & 2.300 (2.505)\\
 6 &  458.1 & 1.107 (0.480) &  0.545 (0.877) &  1.803 (1.598) & 2.080 (2.431)\\
 7 &  602.0 & 1.230 (0.557) & -0.103 (0.923) &  3.783 (1.468) & 1.729 (2.480)\\
 8 &  804.7 & 0.719 (0.585) &  1.555 (0.998) &  2.090 (1.580) & -3.246 (2.921)\\
 9 & 1151.3 & 1.821 (0.674) &  2.748 (1.113) & -4.029 (2.099) & -8.808 (3.388)\\
10 &$\infty$& 4.640 (0.711) &  4.428 (1.200) & -7.147 (2.426) & -10.519 (3.685)\\ \hline
\end{tabular}
\end{center}
\end{table}

We see that the covariates, on the whole, do indeed appear to have an effect on the survival times of patients with leukemia. Age has the most pronounced overall effect, with all of the posterior means of the coefficients being positive. Deprivation score and white blood cell count have positive posterior means for all but the later time periods, indicating that they too could be associated with the hazard. 

The negative posterior means in the later intervals could be an indication that the effects of these covariates decrease over time.  This could be related to the phenomenon of dynamic selection of the population. For example, if, in the later intervals, few older patients are still observed then this could be the reason for a smaller effect of age at later time-points. In our case the mean age of patients in each interval are (60.7, 57.1, 54.5, 53.5, 52.2, 51.1, 50.6, 49.0, 48.5, 48.3) and so dynamic selection may explain the smaller effects in later intervals.

\subsection{Comparison with full Bayesian analysis}

\begin{figure}
\begin{center}

\resizebox{\textwidth}{!}{\includegraphics{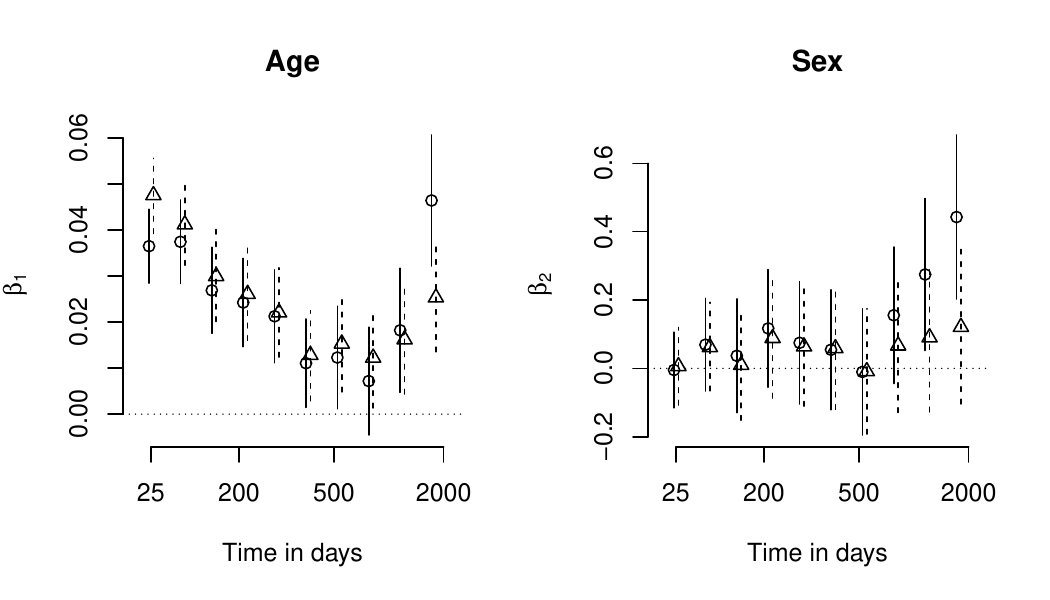}}

\end{center}
\caption{The effects of age and sex on the hazard functions of patients with leukemia. The posterior means for the Bayes linear Bayes method are given by circles and those for the full Bayesian method with triangles. Both are plotted with posterior intervals.}
\label{Bayes}
\end{figure}

We can compare the results of our Bayes linear Bayes analysis with a full Bayesian approach. The full Bayesian model we use for the comparison has piecewise constant hazards over the same intervals as the Bayes linear Bayes  model. The prior first and second moments of the coefficients are the same as for the Bayes linear Bayes model but now we give the parameters a multivariate normal prior distribution. We computed the posterior distribution using a Gibbs sampler with two parallel chains. The first 2000 iterations were discarded and, after this, convergence appeared satisfactory although mixing was not particularly good. The posterior distribution was  calculated using a further 10000 iterations. This took approximately twenty minutes to run on a desktop computer in comparison with about two seconds for the Bayes linear Bayes model. The posterior means and 95$\%$ symmetric posterior intervals for $ \beta_{j,1}$ and $\beta_{j,2}$ are given in Figure \ref{Bayes} alongside the means and $\pm 2$ standard deviations intervals previously given for the Bayes linear Bayes method.

We see that, in most cases, the results using the two methods are similar, but not always. In particular, in the last time period, the posterior means using the full Bayes method are noticeably less than those obtained using the Bayes linear Bayes analysis. Considering Figure \ref{fig:compare2}, we can conclude that the largest discrepancies between the full-Bayes and Bayes linear Bayes results are likely to occur when $\alpha$ is small and the patient dies in the interval in question. In fact, in our example, $\alpha$ is always small. In the case of the log mode method, we have $\alpha_{i,j}=q_{i,j}^{-1}$ where $q_{i,j}$ is the variance of $\eta_{i,j}$. From Table \ref{tab:KJW} we can see that even a baseline patient would have $q_{i,j}=0.64+0.1225=0.7625$ giving $\alpha_{i,j}=1.31$. Other patients will have larger values of $q_{i,j}$ and therefore smaller $\alpha_{i,j}$. A prior variance of 0.8 for a log hazard is really quite large since a $\pm 2$ standard deviation interval for $\eta_{i,j}$ would correspond to an interval for the hazard $\lambda_{i,j}$ with a ratio of 35.8 between the upper and lower limits. Thus, where possible, the use of less diffuse and realistically informative priors is likely to lead to results which are closer to those of the corresponding full Bayes analysis.

Table \ref{post:diff} shows the differences between the full Bayes posterior means and the Bayes linear Bayes posterior means, standardised by dividing by the full Bayes posterior standard deviation. Positive differences imply that the Bayesian method gives larger values for the posterior mean. For comparison, the differences between the full Bayes posterior means and the prior means divided by the prior standard deviations are shown in brackets.

\begin{table}
\caption{Standardised differences between the posterior means, full Bayes minus Bayes linear Bayes, for each of the parameters in each interval.  Standardised differences between the full Bayes posterior means and prior means are shown in brackets.}
\label{post:diff}

\begin{center}
\begin{tabular}{rrrrrrrrr}\hline
Interval $j$ & \multicolumn{2}{c}{$\beta_{j,1}$} & \multicolumn{2}{c}{$\beta_{j,2}$} & \multicolumn{2}{c}{$\beta_{j,3}$} & \multicolumn{2}{c}{$\beta_{j,4}$} \\ \hline
1   &    2.63   &     (1.38)    &     0.19    &     (0.02)   &   -3.72   &   (-0.14)  &    1.52    &     (0.58) \\
2   &    0.84   &     (1.06)    &    -0.11    &     (0.18)   &   -1.60   &   (-0.58)  &    0.19    &     (0.58) \\
3   &    0.61   &     (0.50)    &    -0.33    &     (0.03)   &   -0.68   &   (-0.54)  &   -0.04    &     (0.21) \\
4   &    0.36   &     (0.30)    &    -0.32    &     (0.25)   &   -0.57   &   (-0.32)  &   -0.55    &     (0.27) \\
5   &    0.16   &     (0.10)    &    -0.12    &     (0.18)   &   -0.41   &   (-0.46)  &   -0.51    &     (0.11) \\
6   &    0.35   &    (-0.36)    &     0.05    &     (0.17)   &   -0.22   &   (-0.70)  &   -0.35    &     (0.12) \\
7   &    0.57   &    (-0.23)    &     0.02    &    (-0.02)   &   -0.58   &   (-0.42)  &   -0.16    &     (0.13) \\
8   &    0.92   &    (-0.39)    &    -0.88    &     (0.19)   &    0.06   &   (-0.56)  &    0.79    &    (-0.10) \\
9   &   -0.32   &    (-0.19)    &    -1.65    &     (0.26)   &    2.00   &   (-0.96)  &    1.76    &    (-0.32) \\
10  &   -3.45   &     (0.27)    &    -2.78    &     (0.35)   &    2.69   &   (-1.20)  &    1.78    &    (-0.47)
\end{tabular}
\end{center}

\end{table}

The differences in the case of $ \beta_{j,1} $ and $\beta_{j,2}$ can be seen here. There are also relatively large discrepancies in the case of $ \beta_{j,3}$, the coefficient of white blood cell count, and in the log baseline hazard $ \beta_{j,0} $ (not shown here), although the general patterns of the results given by the two methods are the same.

This was a severe test of the Bayes linear kinematic approach since each observation on an individual in an interval typically contributed little information and thus had relatively little effect on the gamma prior distribution for the corresponding hazard. These gamma prior distributions had small shape parameters $ \alpha$ and Figure \ref{fig:compare2} shows that this leads to the greatest difference from the full Bayes analysis. An observed death would increase $ \alpha $ by 1 while a censoring or the observation that the individual survived the interval would not increase $ \alpha $ at all. In the Bayes linear kinematic approach, each of these $\alpha $ values corresponding to observations would only be changed, if at all, by its own observation and subsequent observations also act on prior gamma distributions with small $ \alpha$.

\section{Simulation}
As a further examination of the behaviour of the Bayes linear Bayes method, we carried out some simulation experiments. We generated simulated data sets and then, in each case, computed posterior means of the parameters using the model and prior specification as used in the leukaemia example. To generate the data sets we used the same number of patients and the same covariate values as in the leukaemia example but we randomly generated survival times and censoring. First we generated a random survival time $t$ for each patient. Then a censoring time $t_{\rm cens}$ for the patient was drawn from an exponential distribution with rate 0.0001. If $t>t_{\rm cens}$ then that patient was labelled as censored at time $t_{\rm cens}$.

\begin{figure}

\begin{center}

\resizebox{\textwidth}{!}{\includegraphics{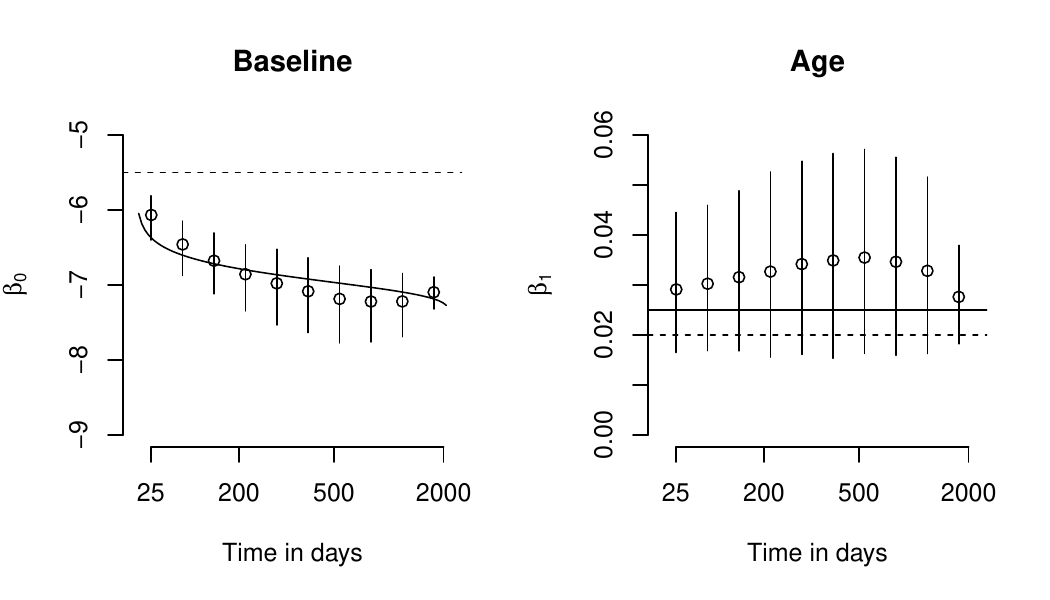}}

\end{center}

\caption{Simulation results where the true distribution is Weibull. Empirical means and 95\% intervals of Bayes linear kinematic posterior means for 2000 simulated samples. True values are shown by the solid line and prior means by the dashed line.}
\label{fig:sim}

\end{figure}

Figure \ref{fig:sim} shows an example of the results. This is based on 2000 simulated data sets. The survival times were drawn from a Weibull distribution with shape parameter 0.8. Thus the baseline hazard decreases over time. The other parameters were set at $\beta_{0}=-5.5,\ \beta_{1}=0.025,\ \beta_{2}=0.1,\ \beta_{3}=0.004,\ \beta_{4}=0.04$. Note that these coefficients did not change over time but the baseline hazard did because of the Weibull shape parameter. Thus, in terms of the piecewise constant hazard model, $\beta_{0}$ changes over time and the other coefficients do not. Figure \ref{fig:sim} shows results for $\beta_{0}$ and $\beta_{1}$. For each time interval, the 2000 posterior means are summarised by their empirical mean and empirical 95\% interval. The true parameter value is shown by a solid line and the prior mean is shown by a dashed line. It can be seen that, in every case, the true value is inside the 95\% interval. Of particular interest is the way that the posterior means of the baseline hazard successfully track the true value even though the Bayes linear Bayes model used for inference contains no information about the form which this would take.

\section{Discussion and Conclusions}
\label{sec:con}
In this paper we have investigated the application of a Bayes linear Bayes model and Bayes linear kinematics to survival analysis using a piecewise constant hazards model with a temporally dependent prior. The approach taken involved fully Bayesian conjugate updates for individual hazards within an interval. These changes in belief were propagated through to hazards for other individuals and intervals using Bayes linear kinematics.

The approach is similar to that of \cite{Gam91}. However, by using Bayes linear kinematics as opposed to standard Bayes linear updating, we are able to produce a solution which is commutative unlike the analysis of \cite{Gam91}. Unlike other published work on Bayes linear Bayes methods, we used a nonlinear link function between the parameters of the observational distributions, in this case the hazards, and the quantities in the Bayes linear structure.

We applied our approach to an example involving patients with acute myeloid leukemia in the North-West of England. There were 1043 patients in the analysis and, due to the analytic nature of all of the updating, posterior means and variances were calculated efficiently. No intensive numerical methods were necessary even though our method allows covariate effects to change over time in a flexible way. We compared the results from our Bayes linear Bayes approach to those from a typical full Bayesian approach. Posteriors in this case were found using Markov chain Monte Carlo methods. The Bayes linear Bayes approach produced solutions far more quickly.
Further research on producing Bayes linear Bayes analyses which more closely match the corresponding full Bayes results is planned.

We have examined a Bayes linear kinematic approach to survival analysis for a piecewise constant hazards model. This does, of course, require a choice of a set of change points. A possible future direction would be to apply our method without arbitrary change-points, in an analysis analogous to a semi-parametric Cox proportional hazards regression. This would be comparable to the reversible jump MCMC work of \cite{Kim07}. 
Frailties \citep[eg.][]{Hend99} can produce an apparent decrease over time in covariate effects but the piecewise constant hazards model is more flexible in the form of changes over time. If required, frailties can be introduced in a straightforward way in our approach by adding a random patient effect $ Z_{i} $ in $ \eta_{i,j}.$

\subsection*{Acknowledgement}
Our thanks go to Professor Rob Henderson for kindly giving us access to the data for the application. We are grateful to referees for helpful comments on earlier drafts of this paper.

\section*{Appendix}

Suppose we observe  the values of unknowns within a Bayes linear structure. Suppose that, as above, we have $ \mathcal{X}= \bigcup_{j=1}^{J} \mathcal{X}_{j}. $  Let  $ {\rm Var}_{0}(X_{j}) = W_{j}. $ In addition, we have other vector random quantities $ Y_{1}, \ldots , Y_{J} $ such that the dimension of  $ Y_{j} $ is $ n_{j} $ and the elements of $ Y_{j} $ are conditionally uncorrelated with those of $ \mathcal{X} \setminus \mathcal{X}_{j} $ given $ X_{j}. $ Moreover we can represent the relationship between $ Y_{j} $ and $ X_{j} $ by writing

\[ Y_{j} = m_{Y,j} + M_{j}(X_{j}-m_{X,j}) + U_{j} \]
where $ m_{X,j} $ and $ m_{Y,j} $ are specified prior mean vectors for $ X_{j} $ and $ Y_{j}, $ $ M_{j} $ is a specified matrix and $ U_{j} $ is a zero-mean random vector with $ {\rm Var}(U_{j}) = V_{j}. $ Hence $ {\rm E}_{0}(Y_{j}) = m_{y,j},\ {\rm Var}_{0}(Y_{j}) = V_{j} + M_{j}W_{j}M_{j}^{\prime} $ and $ {\rm Cov}_{0}(Y_{j},X_{j})=M_{j}W_{j}. $

Let the dimension of $ X $ be $ n_{X}$, let $ n_{Y} = \sum_{j=1}^{J} n_{j} $ and let $ M $ be a $ n_{Y} \times n_{X} $ matrix consisting of zeroes except that, in rows $ 1 + \sum_{j=1}^{k-1} n_{j} $ to $ \sum_{j=1}^{k} n_{j} $  we have the elements of the rows of $ M_{j} $ placed in the columns corresponding to the elements  of $ \mathcal{X}_{j}. $  Let $ Y = (Y_{1}^{\prime}, \ldots , Y_{J}^{\prime})^{\prime},\ m_{Y} = (m_{Y,1}^{\prime}, \ldots , m_{Y,J}^{\prime})^{\prime},\ m_{X} = {\rm E}_{0}(X),\ U=(U_{1}^{\prime}, \ldots , U_{J}^{\prime})^{\prime} $ and let $ {\rm Var}_{0}(U) = V_{U}, $ which is block-diagonal, and $ {\rm Var}_{0}(X)=V_{X}. $ Then $ {\rm Var}_{0}(Y) = M V_{X} M^{\prime} + V_{U} $ and $ {\rm Cov}_{0}(X,Y)=V_{X}M^{\prime} $ so, given an observation $ y $ of the whole of $ Y, $ our adjusted expectation and variance are
\begin{eqnarray*}
{\rm E}(X \mid Y) & = & m_{X} + V_{X} M^{\prime} (M V_{X} M^{\prime} + V_{U} )^{-1} (y - m_{Y}), \\
{\rm Var}(X \mid Y) & = & V_{X} - V_{X} M^{\prime} (M V_{X} M^{\prime} + V_{U})^{-1} M V_{X}.
\end{eqnarray*}

It is easily verified that the precision is
\[ {\rm P}(X \mid Y) = {\rm Var}^{-1}(X \mid Y) = P_{X} + M^{\prime} P_{U} M \]
where $ P_{X} = V_{X}^{-1} $ and $ P_{U}=V_{U}^{-1} $ (where the necessary inverses exist). Furthermore, because of the block-diagonal structure of $ V_{U} $ and $ P_{U}, $
\begin{equation}
{\rm P}(X \mid Y) = P_{X} + \sum_{j=1}^{J} \tilde{M}_{j}^{\prime} P_{j} \tilde{M}_{j}, \label{eq:three}
\end{equation}
where $ P_{j} = V_{j}^{-1} $ and $ \tilde{M}_{j} $ is a $ n_{j} \times  n_{X} $ matrix composed of rows $ 1 + \sum_{j=1}^{k-1} n_{j} $ to $ \sum_{j=1}^{k} n_{j} $  of $ M. $

Similarly

\[ {\rm E}(X \mid Y) = {\rm P}^{-1}(X \mid Y) \{ P_{X}m_{X} + M^{\prime} P_{U}(y-m_{Y})\} \]
and, because of the structure of $ M $ and $ P_{U}, $
\begin{equation}
{\rm E}(X \mid Y) = {\rm P}^{-1}(X \mid Y) \{ P_{X}m_{X} + \sum_{j=1}^{J}\tilde{M}_{j}^{\prime} P_{j}(y_{j}-m_{Y,j})\} \label{eq:four}.
\end{equation}

Now consider what happens if we only observe $ Y_{j}=y_{j}. $ The corresponding results are
\begin{equation}
{\rm P}(X \mid Y_{j}) = P_{X} + \tilde{M}_{j}^{\prime} P_{j} \tilde{M}_{j} \label{eq:five}
\end{equation}
and
\begin{equation}
{\rm E}(X \mid Y_{j}) = {\rm P}^{-1}(X \mid Y_{j}) \{ P_{X} m_{X} + \tilde{M}_{j}^{\prime}P_{j}(y_{j}-m_{Y,j})\}. \label{eq:six}
\end{equation}

From (\ref{eq:three}) and (\ref{eq:five}) we see that
\begin{equation}
{\rm P}(X \mid Y) = \sum_{j=1}^{J} {\rm P}(X \mid Y_{j}) - (J-1) P_{X}. \label{eq:seven}
\end{equation}

From (\ref{eq:four}) and (\ref{eq:six}) we see that
\begin{eqnarray}
{\rm P}(X \mid Y){\rm E}(X \mid Y) & = &  P_{X} m_{X} + \sum_{j=1}^{J} \tilde{M}_{j}^{\prime} P_{j} (y_{j}-m_{Y,j}) \nonumber \\
                                 & = & \sum_{j=1}^{J} \{ P_{X}m_{X} + \tilde{M}_{j}^{\prime} P_{j}(y_{j}-m_{Y,j})\} - (J-1) P_{X}m_{X} \nonumber \\
 & = & \sum_{j=1}^{J} {\rm P}(X \mid Y_{j}) {\rm P}^{-1}(X \mid Y_{j}) \{ P_{X}m_{X} + \tilde{M}_{j}^{\prime} P_{j} (y_{j}-m_{Y,j})\} \nonumber \\
 &   & - (J-1) P_{X} m_{X} \nonumber \\
 & = & \sum_{j=1}^{J} {\rm P}(X \mid Y_{j}) {\rm E}(X \mid Y_{j}) - (J-1) P_{X} m_{X}. \label{eq:eight}
\end{eqnarray}

Now we return to the more general situation of Bayes linear kinematics and suppose that, instead of observing $ Y_{1}, \ldots , Y_{J}, $ which would cause us to adjust our beliefs about $ X, $ we gain other pieces of information, $ D_{1}, \ldots , D_{J}, $ where receiving $ D_{j} $ causes us to revise our beliefs about $ X_{j} $ directly but only affects our beliefs about $ \mathcal{X} \setminus \mathcal{X}_{j} $ through this effect on $ X_{j}. $

\cite{Gol04} derived the conditions under which a commutative Bayes linear kinematic update exists and under which this update is unique. When a unique commutative update exists, it is given by replacing $ Y_{j} $ and $ Y $ with $ D_{j} $ and $ D $ in (\ref{eq:seven}) and (\ref{eq:eight}), giving (\ref{eq:ten}) and (\ref{eq:eleven}).

\bibliographystyle{Chicago} 
\bibliography{references}

\end{document}